\documentclass[a4paper,11pt]{article}
\pdfoutput=1 % if your are submitting a pdflatex (i.e. if you have
\usepackage{jcappub} % for details on the use of the package, please
\usepackage[T1]{fontenc} % if needed

\usepackage{xcolor}
\usepackage{graphicx,color,bm}
\usepackage{amsmath,amssymb}
\usepackage{epstopdf}
\usepackage{ulem}
\usepackage{array}
\usepackage{verbatim}

\usepackage[utf8]{inputenc}
\usepackage{rotating}
\usepackage{doi}
\usepackage{hyperref}

\newcommand{\be}{\begin{equation}}
\newcommand{\ee}{\end{equation}}
\newcommand{\ben}{\begin{equation*}}
\newcommand{\een}{\end{equation*}}
\newcommand{\bea}{\begin{eqnarray}}
\newcommand{\eea}{\end{eqnarray}}
\newcommand{\bean}{\begin{eqnarray*}}
\newcommand{\eean}{\end{eqnarray*}}

\newcommand{\Mpc}{\ensuremath{\,\mathrm{Mpc}}}

\newcommand {\black} {\color{black}}

\newcommand{\sm}{{standard model }}
\usepackage{hyperref}
\newcolumntype{K}[1]{>{\centering\arraybackslash}m{#1}}

\title{ Decaying warm dark matter  and structure formation
}
\author[a]{Jui-Lin Kuo}
\author[b]{Massimiliano Lattanzi}
\author[a,c,d]{Kingman Cheung}
\author[e]{Jos\'e W.~F. Valle}
\affiliation[a]{Department of Physics, National Tsing Hua University, Hsinchu, Taiwan,}
\affiliation[b]{Istituto Nazionale di Fisica Nucleare, sezione di Ferrara, Polo Scientifico e Tecnologico, Edificio C, Via Saragat 1, I-44122 Ferrara, Italy}
\affiliation[c]{Physics Division, National Center for Theoretical Sciences, Hsinchu, Taiwan,}
\affiliation[d]{Division of Quantum Phases and Devices, School of Physics, Konkuk University, Seoul 143-701, Republic of Korea}
\affiliation[e]{AHEP Group, Instituto de F\'{i}sica Corpuscular, C.S.I.C./Universitat de Val\'{e}ncia,
Campus de Paterna, Apartado 22085, E-46071 Val\'{e}ncia, Spain}

\emailAdd{juilinkuo@gapp.nthu.edu.tw}
\emailAdd{lattanzi@fe.infn.it}
\emailAdd{cheung@phys.nthu.edu.tw}
\emailAdd{valle@ific.uv.es}

\abstract{

  We examine the cosmology of warm dark matter (WDM), both stable and
  decaying, from the point of view of structure formation.
  We compare the matter power spectrum associated to WDM masses of
  $1.5\,\mathrm{keV}$ and $0.158\,\mathrm{keV}$, with that expected
  for the stable cold dark matter $\Lambda$CDM$\equiv$SCDM paradigm,
  taken as our reference model.
  We scrutinize the effects associated to the warm nature of dark
  matter, as well as the fact that it decays.
  The decaying warm dark matter (DWDM) scenario is well-motivated,
  emerging in a broad class of particle physics theories where
  neutrino masses arise from the spontaneous breaking of a continuous
  global lepton number symmetry.
  The majoron arises as a Nambu-Goldstone boson, and picks up a mass
  from gravitational effects, that explicitly violate global
  symmetries. The majoron necessarily decays to neutrinos, with an
  amplitude proportional to their tiny mass, which typically gives it
  cosmologically long lifetimes.
  Using N-body simulations we show that our DWDM
  picture leads to a viable alternative to the $\Lambda$CDM scenario, with
  predictions that can differ substantially on small scales.
}

\begin{document}
%\date{\today}
\maketitle
\flushbottom

\section{Introduction}
\label{Sec:Intro}

So far we have failed to identify the nature of what makes up most of
the matter present in the Universe, only a small fraction of which is
the baryonic stuff found in  stellar objects and intergalactic medium. 
The existence of a ``dark matter" component on all scales is inferred
mainly from the gravitational effect it seems to have on visible
matter.
No particle of the \sm can play the role of dark matter, hence it must
be new physics.  For several decades already, there seems to be a
consensus that dark matter must be collisionless yet, to date, its
detailed nature remains a mystery~\cite{Bertone2005279}.
%%%

On the other hand, the discovery of neutrino
oscillations~\cite{Kajita:2016cak,McDonald:2016ixn} indicates the need
for nonzero neutrino masses.
However, underpinning the detailed properties of neutrinos and the
ultimate origin of their mass poses another great challenge for the
\sm of particle physics~\cite{Valle:2015pba}.

%%%%%%%

A tantalizing possibility is that cosmological dark matter is deeply
related to the generation of neutrino masses~\cite{Lattanzi:2014mia}.
For example, dark matter could be a messenger particle associated to
the neutrino mass
generation~\cite{Ma:2006km,Hirsch:2013ola,Merle:2016scw,Bonilla:2016diq}.
Its stability could also reflect a fundamental property of neutrinos,
such as its possible Dirac nature~\cite{Chulia:2016ngi}.
Or it could follow, for example, as a remnant of the symmetry which
accounts for the peculiar pattern of neutrino mixing angles indicated
by the oscillation experiments~\cite{Hirsch:2010ru,Boucenna:2012qb}.
In both cases dark matter would be a stable weakly interacting massive
particle (WIMP).
There are, however, many well-motivated alternatives to WIMP dark
matter. The associated dark matter candidates need not be strictly
stable, while providing viable cosmology.
For example, a decaying gravitino~\cite{Restrepo:2011rj,Choi:2009ng}
provides an attractive scenario for decaying dark matter related to
neutrino physics.

Here we focus on the possibility that the majoron $J$ plays the role of
decaying dark matter. 
This has a two-fold motivation.
Theoretically, the majoron is a very broad concept, emerging as a
Nambu-Goldstone boson in any theory where neutrino masses arise
from the spontaneous breaking of a continuous global symmetry, such as
lepton number~\cite{chikashige:1981ui,Schechter:1981cv}.
 On the other hand, as an alternative to the $\Lambda$CDM
  paradigm, the majoron picture may have the right properties to
  address some potential drawbacks of the standard scenario, such as
  the ``small scale crisis'' which can be alleviated by the warm
  nature of the majoron~\cite{Weinberg:2013aya,Bullock:2017xww}.

The majoron is assumed to acquire a mass $m_J$ through gravitational
instanton effects that explicitly violate global
symmetries~\cite{coleman:1988tj}.
The value $m_J$ of the majoron mass can not be computed by theory.  A
particularly interesting range for the mass is the keV range.
Such keV majoron has been suggested by Berezinsky and Valle (BV) as a
viable decaying dark matter candidate~\cite{Berezinsky:1993fm}.
On general theoretical grounds, the massive majoron is necessarily
unstable, as it couples to neutrinos, with a strength proportional to
their tiny mass~\cite{Schechter:1981cv}.
In order for the majoron to be the dark matter, it must be
cosmologically long-lived, i.e. its lifetime $\tau_J$ should be of the
order of the age of the Universe
$t_0 =13.8\,\mathrm{Gyr}\simeq 4\times10^{17}\,\mathrm{s}$, or larger,
$\tau_J \gtrsim t_0$.
In fact, it has been shown that cosmic microwave
background (CMB) data places a stronger requirement on the majoron
decay rate~\cite{Lattanzi:2007ux}, in order to avoid producing too
much fluctuation power on the largest CMB scales, following the decay
of majoron to neutrinos and the subsequent modifications to the
cosmological gravitational potentials.
In the framework of a simple one-parameter extension of the standard
$\Lambda$CDM model, one finds $\tau_J > 50\,\mathrm{Gyr}$ using WMAP9
data~\cite{Lattanzi:2013uza}.  The limit tightens to
$\tau_J > 160\,\mathrm{Gyr}$ when Planck 2013 data and large-scale
structures linear data from WiggleZ and BOSS are taken into
account\footnote{Note however, that this last limit is obtained
  assuming a model with primordial tensor modes, motivated at the time
  by the BICEP2 claim, so the two limits cannot be directly
  compared.}~\cite{Audren:2014bca}.  Here we adopt the most
conservative limit as our reference choice when discussing DWDM.

Indeed, the massive majoron dark matter model has been shown to be
consistent with CMB data for interesting choices of the relevant
parameters~\cite{Lattanzi:2007ux,Lattanzi:2013uza,Audren:2014bca}.
If the majoron was in thermal equilibrium with the plasma in the early
Universe and decoupled at some later stage, a mass
$m_J = \mathcal{O}(\mathrm{keV})$ would produce the right dark matter
abundance~\cite{Berezinsky:1993fm}. Moreover, a thermal particle with
keV mass would be a WDM candidate.  For a thermal majoron
that decoupled when all the degrees of freedom of the standard model
were still excited, measurements of CMB anisotropies yield the
following constraints \cite{Lattanzi:2013uza}:
$m_J=(0.158\pm0.007)\,\mathrm{keV}$ (68\% C.L.) and
$\tau_J>50\,\mathrm{Gyr}$ (95\% C.L). We note that this value of the
mass is in tension with constraints coming from observations of the
Ly-$\alpha$
forest~\cite{Narayanan:2000tp,Baur:2015jsy,Irsic:2017ixq}. 
There is the alternative possibility that the majoron was never in
thermal equilibrium with the other species in the cosmological plasma. 
For this reason\footnote{In fact, there are no model-independent limits on the majoron mass.}, we
will treat the mass of the majoron as a free parameter and consider different values for it in our
study.

Since the coupling of the majoron to neutrinos $g_\nu$ is proportional
to the neutrino mass~\cite{Schechter:1981cv}, the decay $J \to \nu\nu$
can naturally have a very long lifetime on cosmological scales.
The CMB constraints for the thermal majoron can be shown to imply
$g_\nu < 5\times 10^{-18}$.
Other decay channels may be present, depending on the model. For
example, in type II seesaw models the majoron can also decay to
photons.
The effective (one-loop suppressed) coupling to $\gamma$'s can be
constrained through X- and $\gamma$-ray observations
\cite{Lattanzi:2013uza,Bazzocchi:2008fh}
However, since this coupling is rather model dependent, we will
disregard the radiative decay channel in what follows. This is, in
practice, equivalent to assume that neutrino masses are generated
through simplest type I seesaw mechanism.

In the present paper we examine the effect of decaying warm dark
matter on non-linear structure formation, so far unexplored in the
literature.
The aim of this paper is exactly to fill this gap, and study the
effect of decaying majoron dark matter on structure formation using
N-body simulations.
In fact, we show that such a DWDM majoron expected within the BV
framework does indeed yield a viable cosmology, which can differ
substantially from that of the standard $\Lambda$CDM paradigm.
This happens for two reasons.  First, due to the warm nature of the
 majoron and second, due to the fact that it decays.
 Our paper is organized as follows. In section~\ref{Sec:simulations} we explain the approach employed in our N-body simulations and demonstrate the convergence of our 
 methodology. In section~\ref{Sec:result} we describe our results while in section~\ref{Sec:baryon} we discuss about the possible impact of various baryonic processes.
 Finally, in section~\ref{sec:conclusion} we draw our conclusions and summarize our results, commenting on their possible implications. Additional discussion on structure formation and the WDM mass allowed by Lyman-alpha forest data is given in the appendix.

\section{The simulations}
\label{Sec:simulations}

\subsection{Methodology}
\label{Subsec:method}

In order to scrutinize the novel features of the DWDM scenario
we perform different cosmological simulations, as listed in
Table.~\ref{Tab:simulation}. We consider DM that is either stable or that decays with a lifetime of 
$50\,\mathrm{Gyr}$, which is the lower limit from the CMB obtained in
Ref.~\cite{Lattanzi:2013uza}. We also consider the case of CDM, and two different
cases of WDM. This makes a total of six N-body cosmological simulations. 
To avoid word cluttering in the following, we use 
abbreviations for these simulations, as given in the Tab.~\ref{Tab:simulation}.

In the CDM simulations, the mass of the DM particle is large
enough to suppress free-streaming on the initial matter power
spectrum. In other words, this is the limit of the DM temperature-to-mass ratio going to zero.
In the DWDM case, we consider two values of the DM mass, namely 
$m_J=0.158~\mathrm{keV}$ and $m_J=1.5\,\mathrm{keV}$. The former value, as mentioned in
Sec.~\ref{Sec:Intro}
, is the one that would give the right relic density for a scalar particle, like the majoron,
that decoupled in the early Universe when all the degrees of freedom of the standard model were present.
The latter value can be realized if the majoron has a nonthermal distribution, or if it is thermal 
but its density is diluted by an additional production of entropy after decoupling (both possibilities were described by an effective parameter
  called $\beta$ in~\cite{Lattanzi:2007ux}). In any case, we will remain
agnostic about the production mechanism, and assume a thermal distribution in all our WDM simulations when generating 
initial conditions (see below). This is also in
view of the fact that even if the majoron provides a neat particle physics motivation for the DWDM scenario,
nevertheless our results are more general, in the sense that they apply independently of the particular nature of DM.

\begin{table}[t]
\begin{center}
\begin{tabular}{p{2.5cm}<{\centering}|p{3.5cm}<{\centering}|p{3.5cm}<{\centering}|p{3.5cm}<{\centering}}
\hline
\hline
Abbreviations & Initial Conditions & Lifetime & WDM mass\\
\hline
SCDM & CDM & $\infty$ & N/A\\ 
DCDM & CDM & $50\,\mathrm{Gyr}$ & N/A\\ 
SWDM-M & WDM & $\infty$ & $1.5\,\mathrm{keV}$\\ 
DWDM-M & WDM & $50\,\mathrm{Gyr}$ & $1.5\,\mathrm{keV}$\\ 
SWDM-m & WDM & $\infty$ & $0.158\,\mathrm{keV}$\\ 
DWDM-m & WDM & $50\,\mathrm{Gyr}$ & $0.158\,\mathrm{keV}$\\ 
\hline
\hline
\end{tabular}
\caption{The abbreviations and features of the simulations we have
  performed in this article. To avoid word cluttering in the following
  we will use these abbreviations.}
\label{Tab:simulation}
\end{center}
\end{table}

The values that we choose for the DM mass are in tension with lower limits obtained
from observations of Ly$-\alpha$ flux-power spectra. For example, the recent analysis
of Ref~\cite{Irsic:2017ixq} finds $m>5.3 \,\mathrm{keV}$ at 95\% CL for a thermal candidate, from a combined analysis of the XQ-100 and HIRES/MIKE data samples. This limit can be relaxed
to $3.5 \,\mathrm{keV}$ by allowing for a non-smooth evolution of the temperature
of the intergalactic medium (IGM). We choose to consider smaller values of the mass
for two reasons. The first one is that the nature of our paper is exploratory, and the main purpose is to study the joint effects of the DM decay and free streaming. A small value of the mass allows us to maximize free-streaming in order to better highlight the interplay between these two effects, taking into account also the computational resources at our disposal. The second reason is that the interpretation of 
Ly$-\alpha$ data is somehow complicated by several factors, like for example the aforementioned dependence on the modeling of the IGM thermal history. For example, Ref~\cite{Garzilli:2018jqh} finds that the Ly$-\alpha$ data can be made consistent with models excluded by other analyses. This, however, does not necessarily imply that thermal DM with the masses considered here can be made consistent with Ly$-\alpha$ observations; a dedicated study would be necessary 
for that purpose. That said we have, in any case, also performed simulations for ``large'' DM mass, $m_J = 5.3 \, \mathrm{keV}$. We found no appreciable difference with the CDM case in the range of scales that we are able to probe within our numerical resolution. The results for that case are given in the appendix.  A future analysis might consider different values of the mass, using larger-resolution simulations, and also a non-thermal spectrum for the DM.

The standard N-body simulation code
\texttt{Gadget2}~\cite{Springel:2005mi} is adopted to perform the
simulations. \texttt{Gadget2} follows the evolution of a
self-gravitating system of collisionless ``particles'', taking into
account the expansion of the Universe. These particles are in fact
macroscopical objects, composed by a large number of DM particles. For
this reason one usually refers to them as ``simulation particles'', as
opposed to actual DM particles. In order to implement the effect of
decay, we include two modifications in the original \texttt{Gadget2}
code, following the approach in
Refs.~\cite{1987ApJ...321...36S,Enqvist:2015ara}, which addressed the
issue of dark matter decays into dark radiation.
Although here we are concerned with dark matter decaying into
relativistic neutrinos, the algorithm of the simulation is similar to
Ref.~\cite{Enqvist:2015ara}.  First of all, the mass of the simulation particle is
reduced by a small amount at each step in the simulation, in order to account for the effect of DM decay.
Therefore, in the simulation the mass of the simulation particles is
altered according to
\begin{equation}
M(t) = M(1-R+R\, e^{-t(z)/\tau_J}),
\label{eq:Mt}
\end{equation}
where $M$ is the initial mass of the simulation particles 
%in the initial condition
, and $R \equiv (\Omega_M - \Omega_b)/\Omega_M$ is the DM
fraction in the matter component, and $\Omega_b$ refers to the
  baryon contribution.
In addition to reducing the simulation particle mass, we also modify the
expansion rate of the universe in accordance with the energy content
at each redshift.
Due to the dark matter decaying into relativistic particles (in the case of the majoron, neutrinos), the expansion history
in the DWDM majoron scenario is different from that of the stable DM
case. The evolution of the dark matter and of the decay products $\rho_{dm}$ and $\rho_{dp}$  are described by 
\begin{equation}
\label{Eq:rho_evolution}
\begin{gathered}
{\dot{\rho}_{dm}}+3\mathcal{H} \rho_{dm} = -\dfrac{a}{\tau_J} \rho_{dm}, \\
{\dot{\rho}_{dp}}+4\mathcal{H} \rho_{dp} = \dfrac{a}{\tau_J} \rho_{dm}, 
\end{gathered}
\end{equation} 
where $\mathcal{H}$ and $a$ are the conformal Hubble parameter and the
scale factor, and the dot represents the derivative with respect to
the conformal time.  Here we assume that the decay products are 
relativistic, so the pre-factor for the Hubble drag term for this component 
in Eq.~\eqref{Eq:rho_evolution} is $4$.  On
the other hand, $\mathcal{H}$ at each redshift is determined by
\begin{equation}
\label{Eq:hubble}
\mathcal{H}^2(z) = \dfrac{ 8\pi G}{3} a^2 (\rho_{dm}(z) + \rho_{b}(z) + \rho_{dp}(z) + \rho_\Lambda(z)),
\end{equation}
where $G$, $\rho_b$, and $\rho_\Lambda$ are the gravitational
constant, the baryon energy density, and the energy density of dark
energy, respectively. We assume that dark energy is in the form of a 
cosmological constant. We also neglect the presence of the 
thermal relic neutrinos produced 
in the early phases of the cosmological evolution, both 
at the background and perturbation level. Note that $\rho_{b}$ and $\rho_\Lambda$ are
unaffected by the energy exchange between DM and the decay products,
hence they evolve as in the standard case 
(i.e., $\rho_b \propto a ^{-3}$ and $\rho_\Lambda = \mathrm{const}$).  
Therefore, given the
initial values for $\rho_{dm}$ and $\rho_{dp}$, we need to numerically
solve Eq.~\eqref{Eq:rho_evolution} in conjunction with
Eq.~\eqref{Eq:hubble} at each timestep, in order to obtain the precise
Hubble parameter describing the expansion of the universe.

For simplicity, in the simulation we neglect the effects of perturbations in the decay products. Indeed, we note that the contribution of the decay products to the energy density is very small, since we consider very long DM lifetimes. Moreover, the decay-produced neutrinos are free-streaming and thus do not cluster, due to their relativistic nature. So the main effect of the decay products is just to reduce the amount of matter that is able to cluster, and this is fully captured by decreasing the mass of each simulation particle as in Eq.~\ref{eq:Mt}. We expect this approximation to break down on the largest scales, above the free-streaming length of the decay products, where these are able to cluster. However, this happens around the horizon scale, which is much larger than the largest scales probed by our simulations, that use a box size of $50 h^{-1} \Mpc$. Moreover, the power spectrum on those scales can be reliably computed using linear theory, if necessary.
As a result, we expect that addional effects related to perturbations in the decay-produced neutrinos will be subtle and not change our results significantly. 

Note that we do not include baryons in our simulation and thus neglect, among others, baryonic feedback processes. The reason again is that, given the scope of our paper, we want to focus on the interplay between DM decay and free streaming. The inclusion of baryonic effects, through hydrodynamic simulations, would be of course mandatory for a rigorous comparison
between the predictions of the ``full-fledged'' DWDM scenario and the observations. We comment, anyway, on the possible effects of baryonic physics in Sec.~\ref{Sec:baryon}.

\subsection{Initial Conditions}
\label{Subsec:IC}
\begin{figure}
\begin{center}
\includegraphics[width=0.45\textwidth]{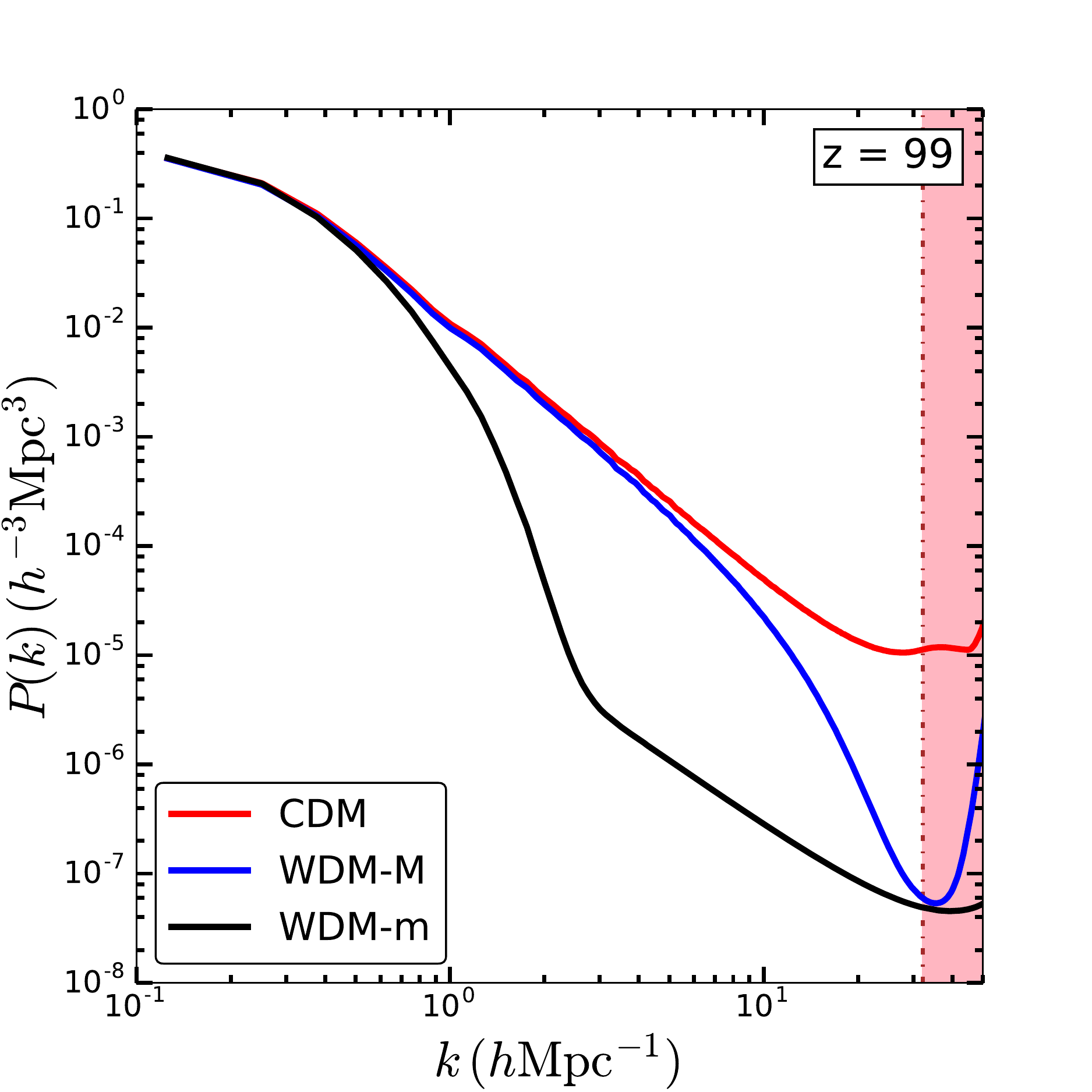}
\includegraphics[width=0.45\textwidth]{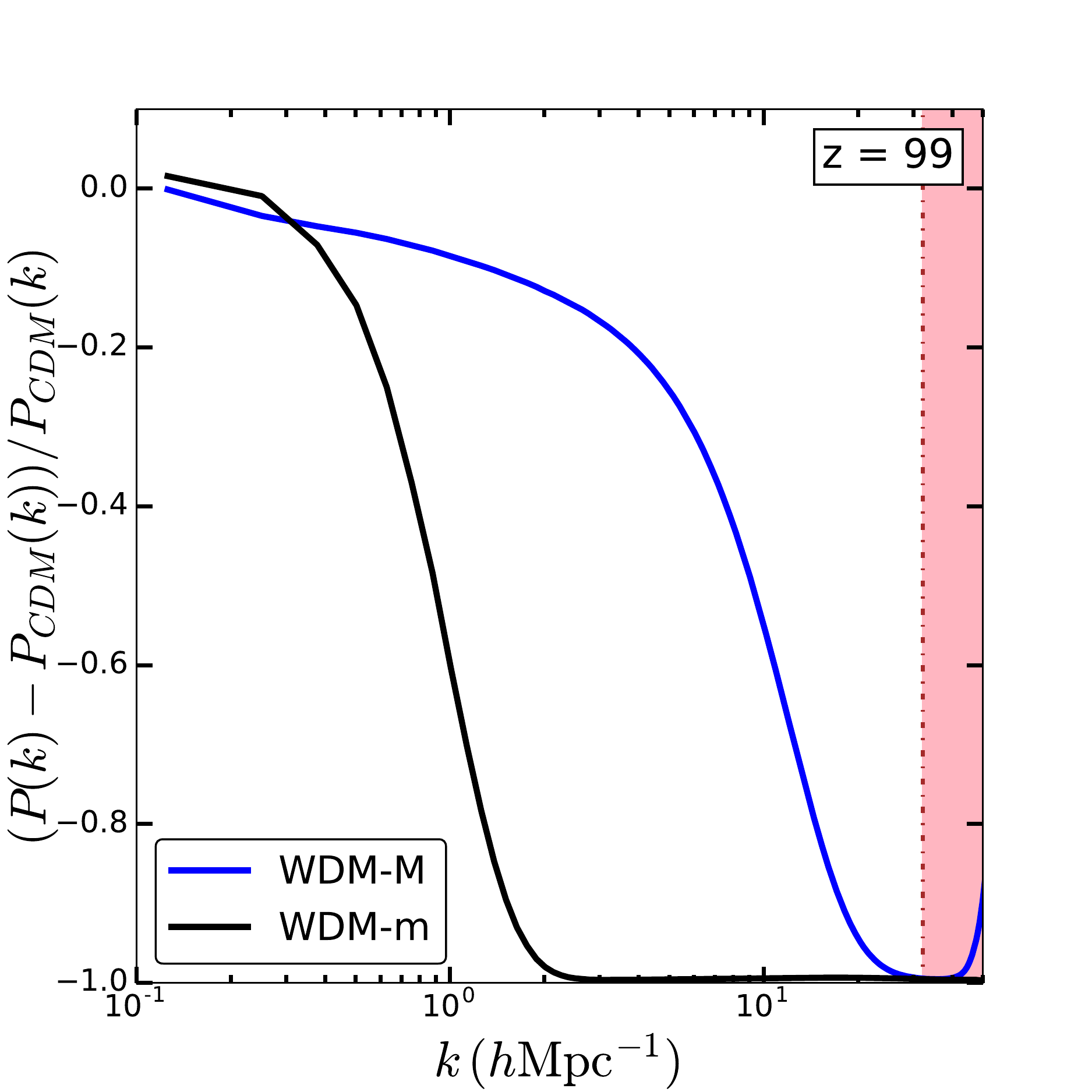}
\caption{Left panel: comparison of the matter power spectrum of initial condition for $\Lambda$CDM (red solid line), WDM with mass $1.5\,\mathrm{keV}$ (blue solid line) and WDM with mass $0.158\,\mathrm{keV}$ (black solid line).  The vertical ``Nyquist''
  band lies above the limit set by the scale of the average size of the simulation particle.  Right panel: 
  Relative difference between the WDM and CDM power spectra, for WDM with mass $1.5\,\mathrm{keV}$ (blue solid line) and
  $0.158\,\mathrm{keV}$ (black solid line). Here the matter power spectra are obtained from the output of \texttt{2LPTic},
  hence the effect of finite numerical resolution is already included.
 The cut-off due to the free-streaming of WDM can be clearly seen.}
\label{Fig:ICPkcompare}
\end{center}
\end{figure}

To generate initial conditions for the N-body simulations,
one uses linear theory to evolve the primordial perturbations 
in $k$ space
up to some redshift deep in the
matter-dominated era, but still early enough for the linear 
predictions to be valid. This is the initial redshift, in our case $z=99$, 
from which the N-body simulations start. Since this initial time 
is well before the DM decay kicks off, the initial power spectra
for the stable and decaying DM case are the same.
We adopted the fitting form of CDM matter power spectrum
$P_\mathrm{CDM}$ given in Ref.~\cite{Eisenstein:1997jh}, 
which is based on the calculation of linear theory, 
to compute the initial power spectrum for CDM initial conditions.
In the WDM scenario, we estimate the power spectrum
at the initial redshift as
\begin{equation}
P_\mathrm{WDM}(k) = T^2_{\mathrm{WDM}}(k)\times  P_\mathrm{CDM}(k) \, ,
\end{equation}
where $T_\mathrm{WDM}(k)$ is the transfer function
given in Ref.~\cite{Bode:2000gq} (where it is called $T_\chi$),
which accounts for the cut-off in
the matter power spectrum due to the free-streaming
effect. The initial
transfer function for thermal WDM can be written as
\begin{equation}
T_{\mathrm{WDM}}(k) = \left(1+(\alpha k)^{2\nu} \right)^{-5/\nu},
\label{eq:wdmtransfer}
\end{equation}
where
$\alpha = 0.048(\Omega_{DM}/0.4)^{0.15} (h/0.65)^{1.3}
(\mathrm{keV}/m_\mathrm{DM})^{1.15} (1.5/g)^{0.29}\,\Mpc$
and $\nu =1.2$. Here $\Omega_{DM}$ is the dark matter energy density,
$m_\mathrm{DM}\equiv m_J$ is the dark matter mass and
$g$ is the effective number of dark matter degrees of freedom
 ($g=1$ for the majoron). Note that $\alpha$ is a 
critical length that determines the cut-off scale in the initial
power spectrum. Using $m_J=0.158\,\mathrm{keV}$ or $1.5\,\mathrm{keV}$,
$g=1$, and the values listed below for the other parameters,
one has that the transfer function reduces the initial fluctuation 
power to a fraction $1/e$ of the corresponding CDM value
at $k\simeq1$ and $17\,h\Mpc^{-1}$, respectively. We can take
these values as rough estimates of the free-streaming wavenumber.

In order to generate the initial condition for the cosmological
simulation we used the \texttt{2LPTic} code~\cite{Crocce:2006ve},
based on the second-order Lagrangian perturbation theory. 
In Fig.~\ref{Fig:ICPkcompare}, we show the initial ($z=99$) power
spectra for CDM and for the two WDM models considered here, given by \texttt{2LPTic}, 
hence numerical limitations are already included~\footnote{The sudden change of slope for the WDM-m scenario around $k=3\, h\Mpc^{-1}$ is due to the presence of shot
noise, which will be discussed in Sec.~\ref{Subsec:convergence}.}.

Note that when we extract initial conditions from the power spectrum,
we use the same
random seed for each pair of stable/decaying DM simulations.
In other words, 
the two simulations of each pair have exactly the same initial conditions. 
The simulations start from redshift $z=99$.  The input cosmological
parameters are: the matter energy density $\Omega_{m}=0.3$, the
cosmological constant energy density $\Omega_{\Lambda}=0.7$, the
baryon energy density $\Omega_{b}=0.04$, the dimensionless Hubble
constant $h=0.7$, the scalar spectral index $n_s=0.96$, and the power
spectrum normalization factor $\sigma_{8}=0.8$. 
For WDM simulations, we input thermal velocities at $z=99$ to the 
simulation particles, consistently with the initial spectrum. 
This has however a negligible effect on
nonlinear structure formation since thermal velocities have
 already decayed out at $z=99$, due to the expansion of the Universe.
We have used $512^3$ simulation particles and
a cube containing these particles with each
side equals to $50\,h^{-1}\Mpc$.
The mass $M_\mathrm{sim}$ of each simulation particle at the initial 
time is $M_\mathrm{sim}\simeq7.8\times 10^{7} \,h^{-1} M_{\odot}$.
  Periodic boundary conditions are
employed in order to avoid boundary effects.  \black

\subsection{Numerical convergence tests}
\label{Subsec:convergence}

In this section we quantify the degree of convergence of our simulations. We do this by considering simulations with different volume and number of particles. In particular, we change the resolution of the simulations at fixed volume, or change the size of the simulation volume at fixed resolution. This will allow us to assess numerical limitations and to define the limit of validity of the results inferred from our simulations.

There are two kinds of numerical limitations. The first is sample variance,  also known as "cosmic variance". In the simulation, the source of sample variance is the finite volume of the simulation and the fact that each simulation only provides a single realization of the underlying statistical distribution of particles. The sample variance prevents us from precisely predicting the density field on large scales~\cite{vanDaalen:2011xb}. We use for our simulations a box size of $50\,h^{-1}\Mpc$, corresponding to a fundamental mode $k \simeq 0.13 \,h\Mpc^{-1}$.
The second kind of numerical limitation is due to the discreteness of the simulation particles, i.e. to the fact that we adopt particles to represent a continuous density field.
The overall resolution limit of the simulations is set by both the box size and the number of particles, and is described by the Nyquist wavenumber $k_\mathrm{Nyq}$
\begin{equation}
k_\mathrm{Nyq} = \pi (N/V)^{1/3}.
\label{eq:nyq}
\end{equation}

Beyond the Nyquist wavenumber, the accuracy of the power spectrum is strongly degraded. For the parameters used in our baseline simulations,
$k_\mathrm{Nyq} \simeq 32 \, h\Mpc^{-1}$.

The finite resolution of the simulations has two consequences. First, non-zero power exists on all scales, called shot noise. The shot noise originates because we adopt particles to represent a continuous density field. The amplitude of the shot noise is independent of the wavenumber $k$ and depends instead on the number of particles in the simulation, and therefore on the resolution~\cite{Colombi:2008dw,vanDaalen:2011xb}. The second consequence is a discreteness peak in the power spectrum at twice the Nyquist limit. This small excess of power is a common feature of all the N-body simulations. This is however more of a problem for WDM simulations than CDM ones, since the former have much less power at small scales, making this numerical artifact more manifest. This causes the well-known spurious halo issues in standard WDM simulations~\cite{Wang:2007he}. We will discuss this effect in more details in Sec.~\ref{Subsec:HMF}.

As anticipated above, in order to test the convergence of our simulations, we compare the results from runs
with different box size and particle resolution. To this purpose, we perform simulations with $N=128^3,\,256^3,\,512^3$
and $L=V^{1/3}=50,\,100\,h^{-1}\Mpc$.
We then compare the resulting power spectra at $z=0$ with that from our baseline run
with $L= 50\,h^{-1}\Mpc$ and $N= 512^3$.
We concentrate on the DWDM-m case, i.e. the one with the smaller mass for the dark matter particle.
This is because this is the case with the stronger suppression of small-scale power, for which
numerical issues at small scales are thus in principle more relevant.
In order to assess the impact of box size, we do the following. We compute the
ratio of the matter power spectra at $z=0$ from the simulations with $\{L,\,N\} = \{50h^{-1}\Mpc,\,128^3\}$ and $\{100h^{-1}\Mpc,\,256^3\}$. The latter has different number of particles to ensure that the two simulations have the same resolution ($k_\mathrm{ny} \simeq 8 h\Mpc^{-1}$ in
both cases) and that we are isolating the effects of the finite simulation volume. The ratio between the spectra should give us a rough measure of the numerical error associated to a finite volume size of $50h^{-1}\Mpc$, at that resolution. We also do the same for the pair of simulations with
$\{L,\,N\} = \{50h^{-1}\Mpc,\,256^3\}$ and $\{100h^{-1}\Mpc,\,512^3\}$ ($k_\mathrm{ny} \simeq 16 h\Mpc^{-1}$) to 
be confident that our results reliably extrapolate to our reference simulation with $\{L,\,N\} = \{50h^{-1}\Mpc,\,512^3\}$ and $k_\mathrm{ny} \simeq 32 h\Mpc^{-1}$. Of course a more direct way would be to perform a simulation with $\{L,\,N\} = \{100h^{-1}\Mpc,\,1024^3\}$, but
we choose not to follow this path due to our limited computational resources.

\begin{figure}
\begin{center}
\includegraphics[width=0.45\textwidth]{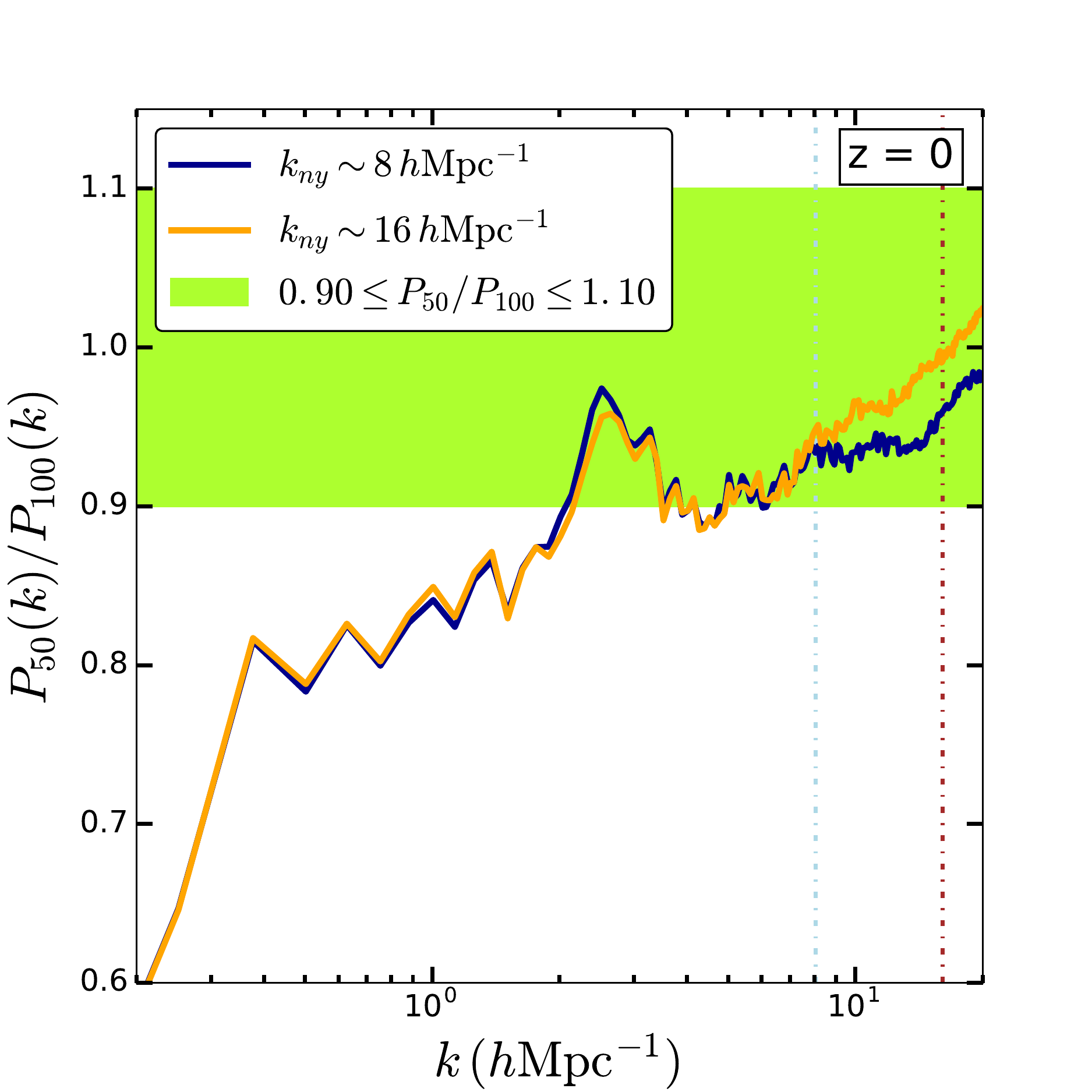}
\caption{ Ratio of power spectra obtained from simulations with box-size $L=50$ and $100 h^{-1}\Mpc$, at fixed resolution.
The blue and yellow curves correspond to $k_\mathrm{ny} \simeq 8$ and $16 h\Mpc^{-1}$, respectively. The green band shows a $10\%$ deviation between the spectra.}
\label{Fig:Pk_ratio_50_100}
\end{center}
\end{figure}

We show the ratio of the spectra computed in this way in Fig.~\ref{Fig:Pk_ratio_50_100}. It is evident how the large-scale power of the simulations does not match due to the cosmic variance. However, we see that for both resolutions, the relative difference
between $L=50$ and $100 h^{-1}\Mpc$ is $10\%$ or better at wavenumbers above $k\simeq 2 h \Mpc^{-1}$. This makes us confident that the same applies at the resolution of our reference simulation.

\begin{figure}
\begin{center}
\includegraphics[width=0.45\textwidth]{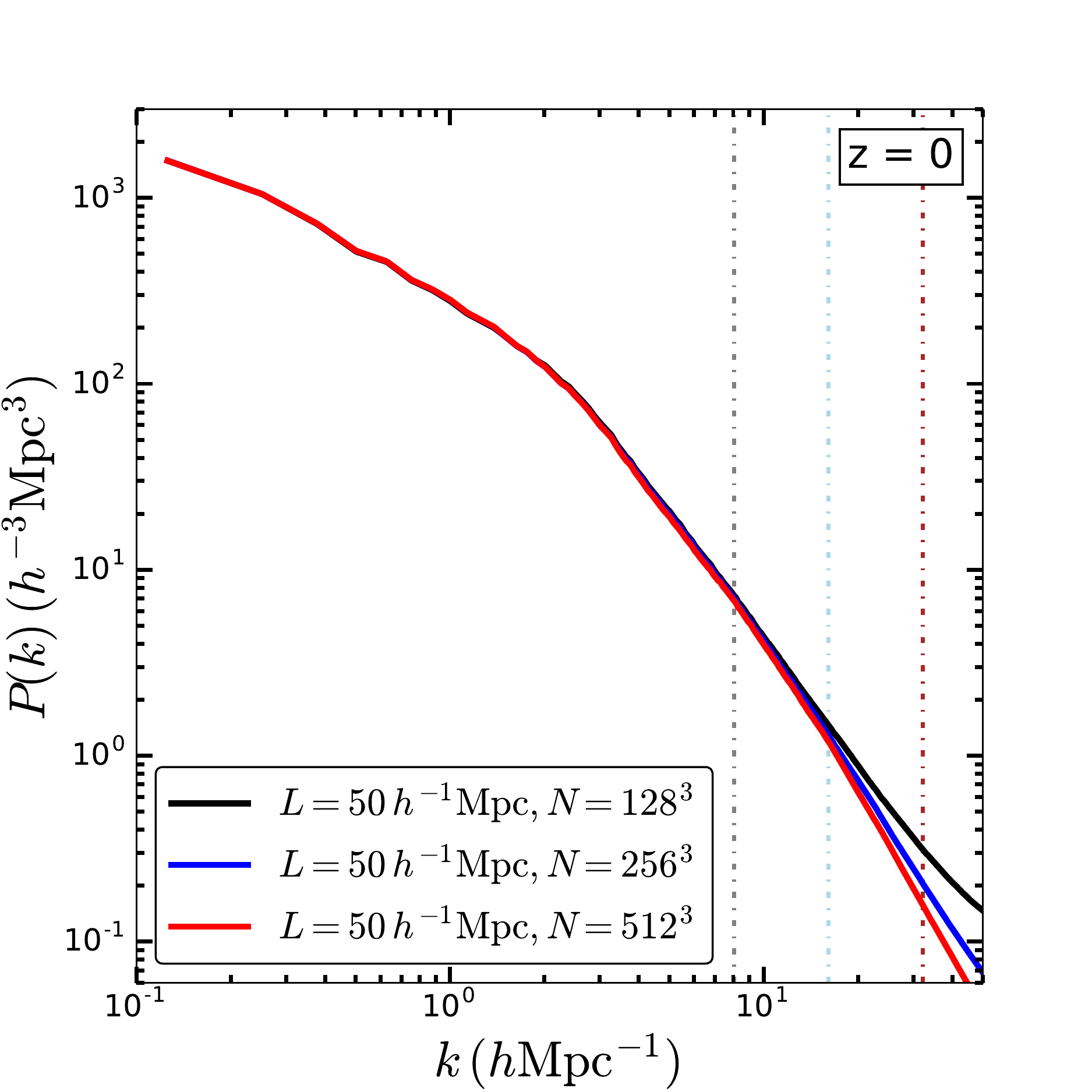}
\includegraphics[width=0.45\textwidth]{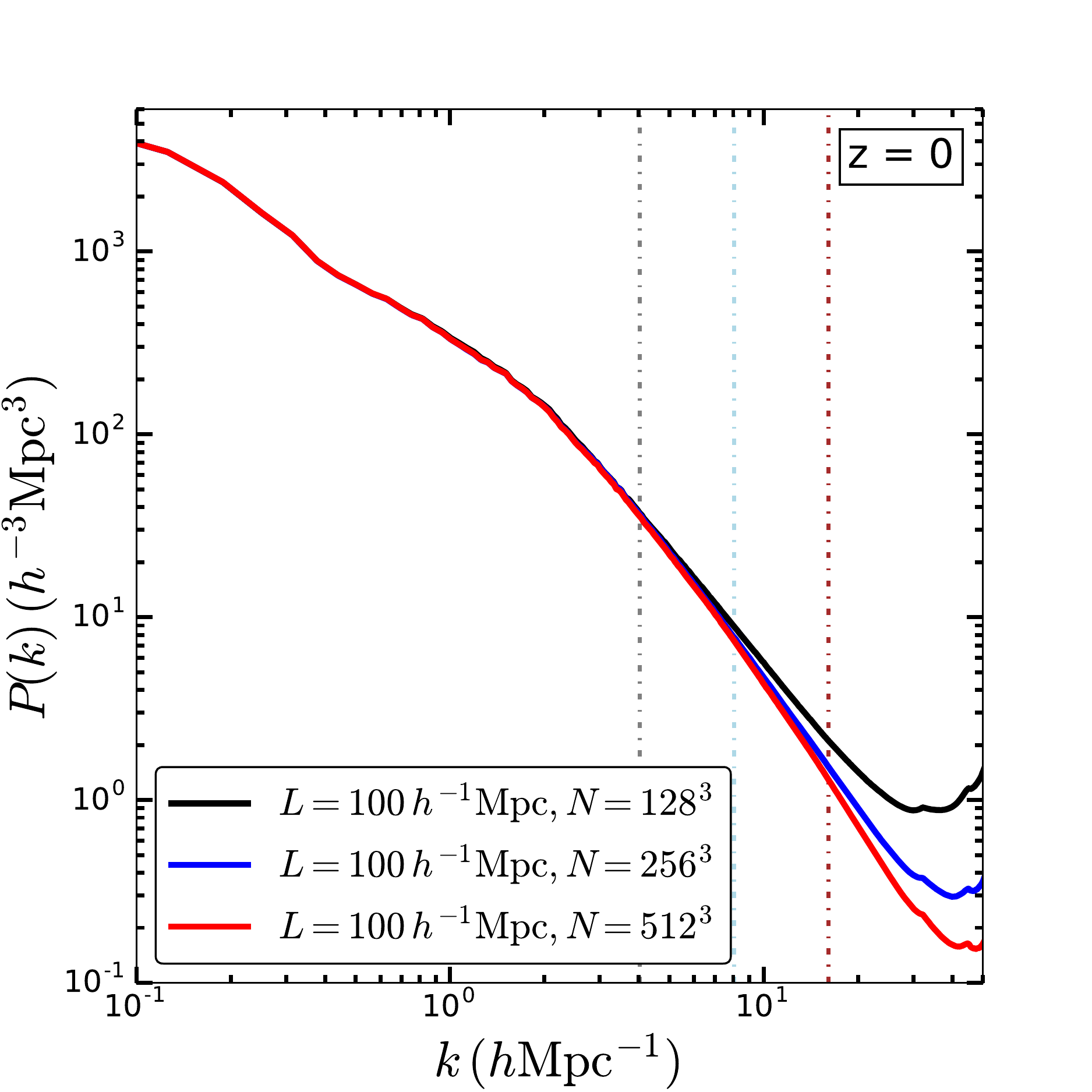}
\caption{The matter power spectrum at $z=0$ of different simulation resolutions with 
$V=50^3\,h^{-3}\Mpc^3$ (left panel) and $V=100^3\,h^{-3}\Mpc^3$ (right panel).
One can see that in almost the entire range of scales the matter power spectra at 
different resolutions converge, all the way up to the Nyquist limit.}
\label{Fig:Pk_z0_compare_dif_scale}
\end{center}
\end{figure}

\begin{figure}
\begin{center}
\includegraphics[width=0.45\textwidth]{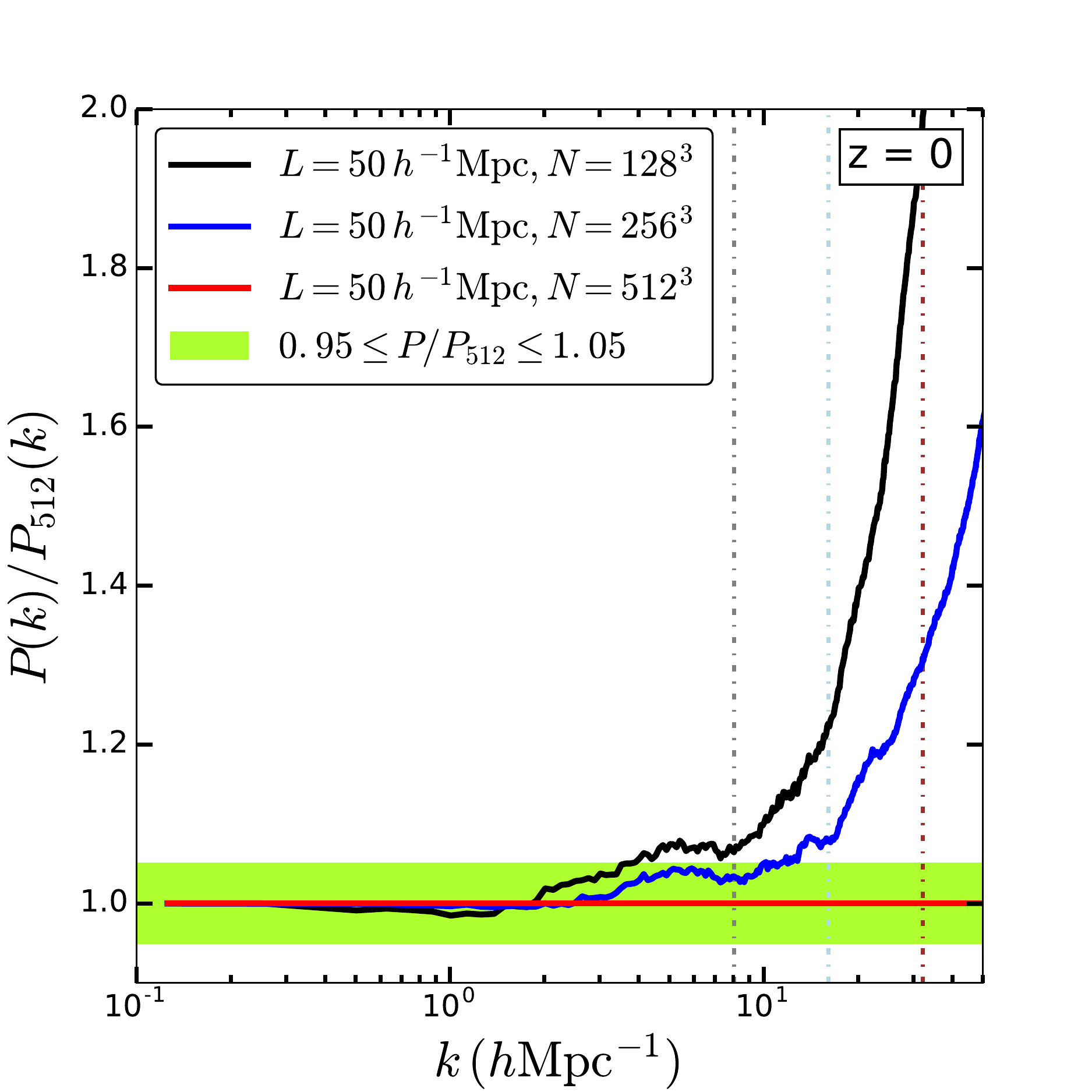}
\includegraphics[width=0.45\textwidth]{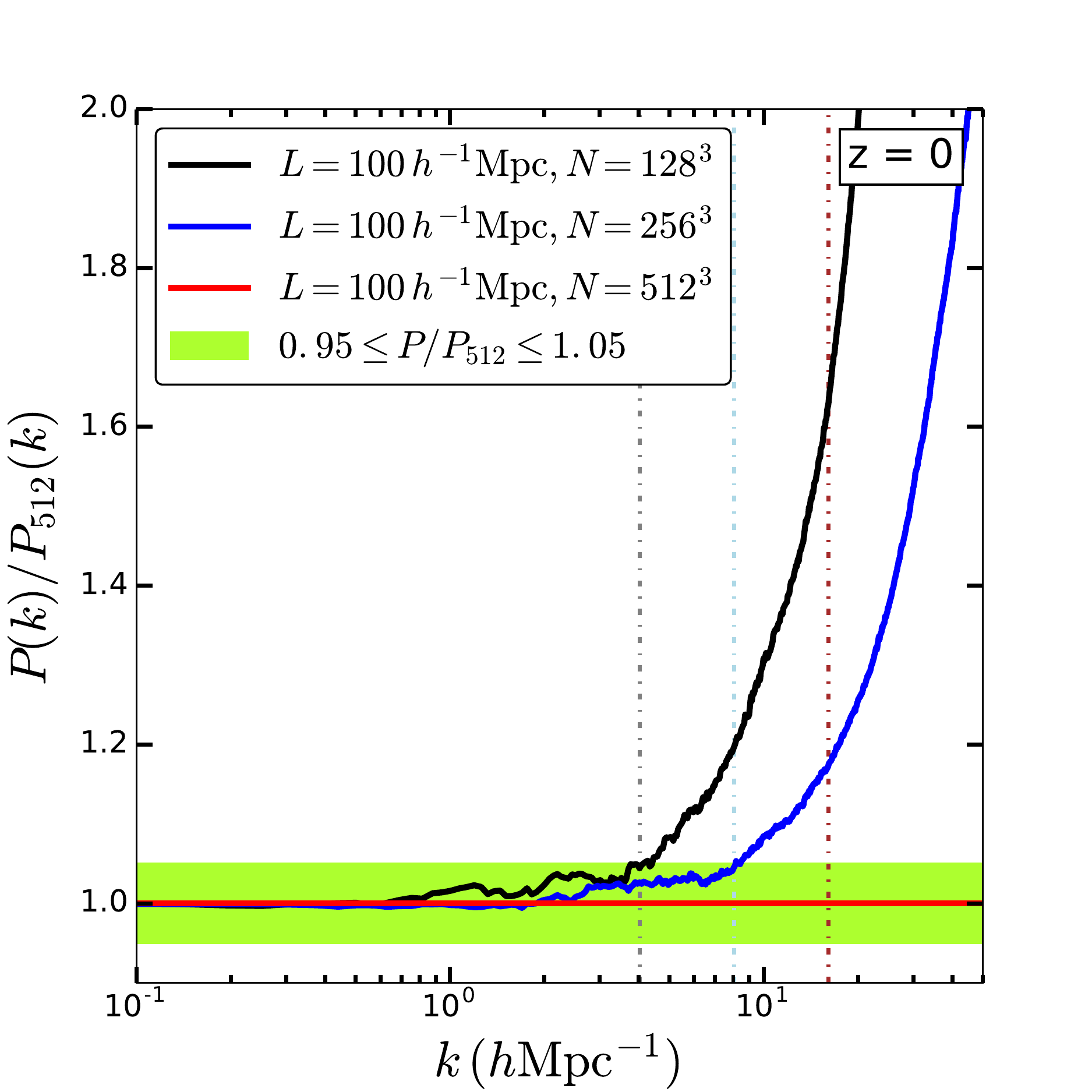}
\caption{Effect of changing the simulation resolution at fixed box size $L$. The solid curves show the ratio
between the matter power spectra at $z=0$ of Fig.~\ref{Fig:Pk_z0_compare_dif_scale}, obtained with the settings for the particle number indicated in the legend, and the spectrum for $N=512^3$ choosen as reference. 
The left (right) panel is for $L=50\,(100) \,h^{-1}\Mpc$.}
\label{Fig:Pk_z0_compare_dif_scale_ratio}
\end{center}
\end{figure}

Then, to study the effect of the finite resolution, in the left panel of Fig.~\ref{Fig:Pk_z0_compare_dif_scale}
we show matter power spectra at $z=0$ from simulations with $50^3\,h^{-3}\Mpc^3$ 
box size and $N^{1/3}=128,\,256,\,512$. Values of $k_\mathrm{Ny}$  for this runs are $8,\,16$ and $32 \, h\,\Mpc^{-1}$. The right panel of the same figure shows the corresponding plot for a $100^3\,h^{-3}\Mpc^3$ box size, with $k_\mathrm{Ny}= 4,\,8$ and $16 \, h\,\Mpc^{-1}$.
On large and intermediate scales, the matter power spectra at different resolutions converge fairly well, starting to deviate beyond the Nyquist wavenumber $k_\mathrm{Nyq}$ of the given resolution. In particular, the excess of power above the Nyquist wavenumber is a manifestation of particle shot noise.
In order to better highlight this effect, we show in the two panels of Fig~\ref{Fig:Pk_z0_compare_dif_scale_ratio} the ratios between each of the spectra and a (third) reference spectrum $P_{512}(k)$ for the $N=512^3$ case, evaluated at $z=0$. It can be seen that the simulations agree to within 5\% or better below the Nyquist wavenumber. In particular, at $k=k_\mathrm{Ny}/2$ the error at $z=0$ is $5.7\%$ for the $128^3$ particles run and $3.4\%$ for the $256^3$ run, for a box size of $50\,h^{-1}\Mpc$. The corresponding numbers for the $100\,h^{-1}\Mpc$ boxsize are $2.5\%$ and $2.6\%$.

From the results presented in this section, it is clear that the parameter set $V=50^3\,h^{-3}\Mpc^3$ and $N=512^3$ provides an adequate benchmark choice for our simulations. In particular, we find that our simulations have $\sim 10\%$ accuracy or better in the wavenumber range $(1 - 20) h\Mpc^{-1}$.

\section{Simulation Results}
\label{Sec:result}

  By comparing the results of our simulations, we can
  infer the effect of DWDM on structure formation. In the
  following, we derive our results through detailed analyses
  of the density field, the matter power spectrum and the halo mass
  function inferred from our N-body simulations.

\subsection{Density Field}

\begin{figure}[t]
\centering
\includegraphics[width=0.31\textwidth]{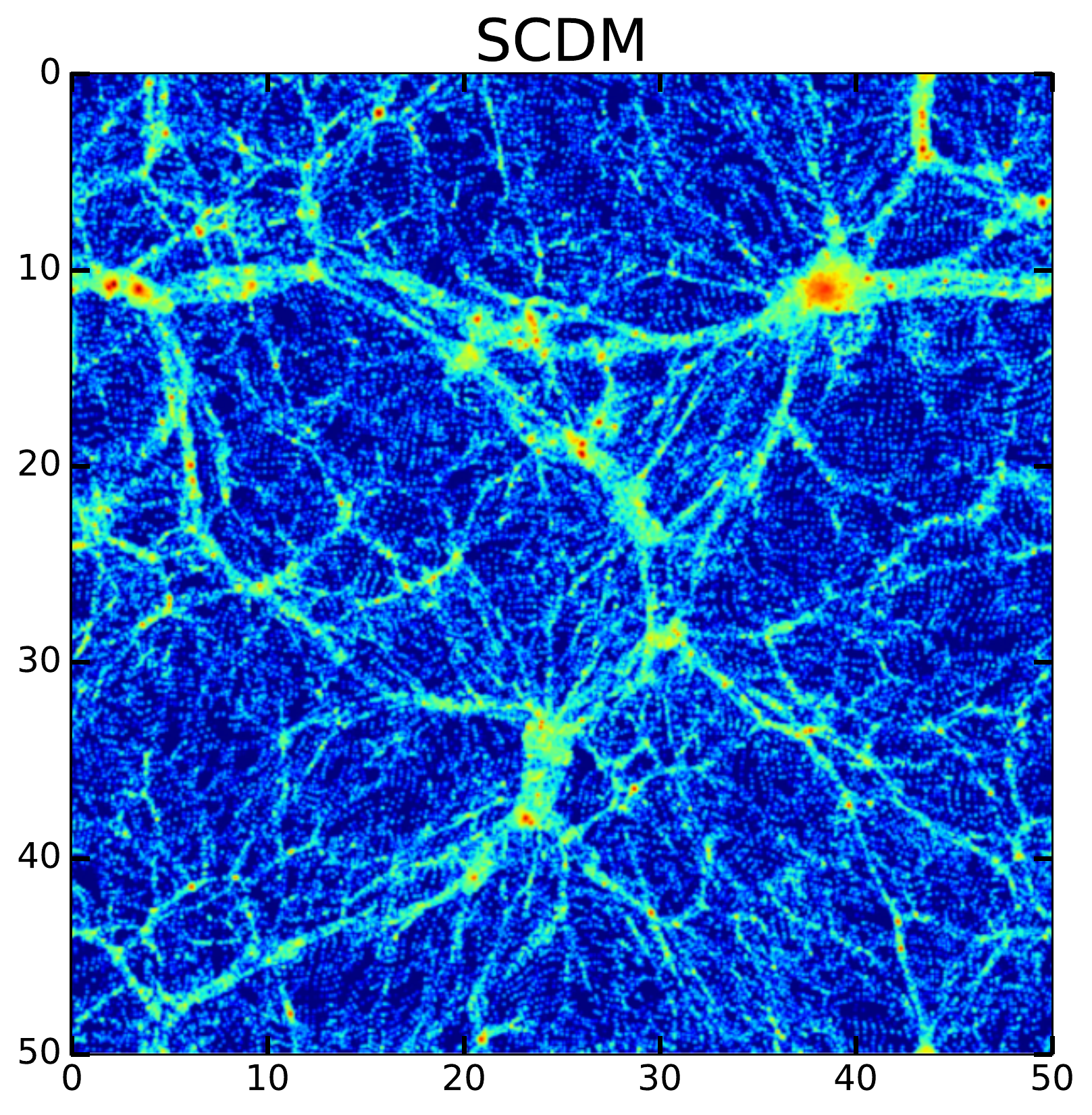}
\includegraphics[width=0.31\textwidth]{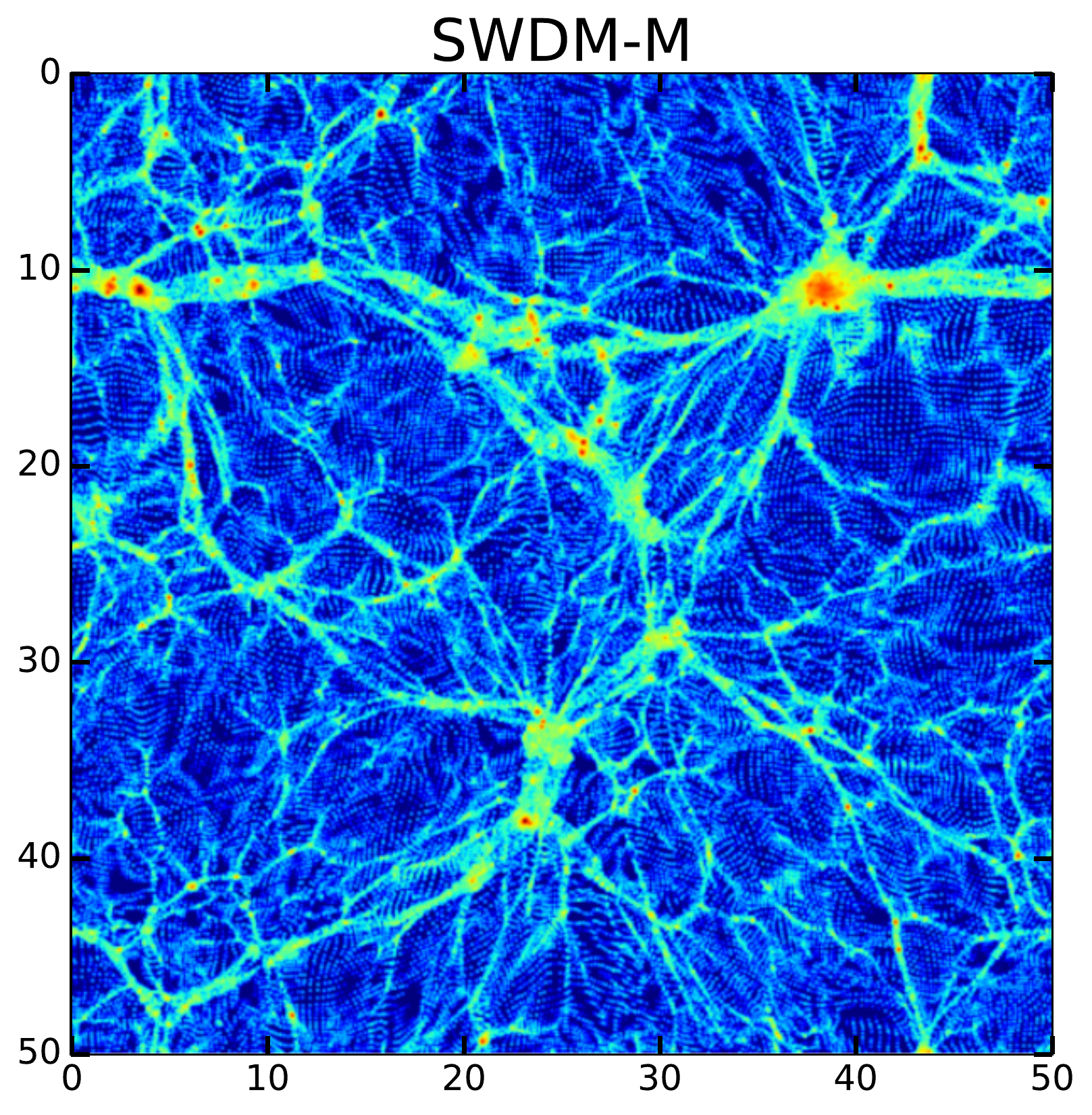}
\includegraphics[width=0.31\textwidth]{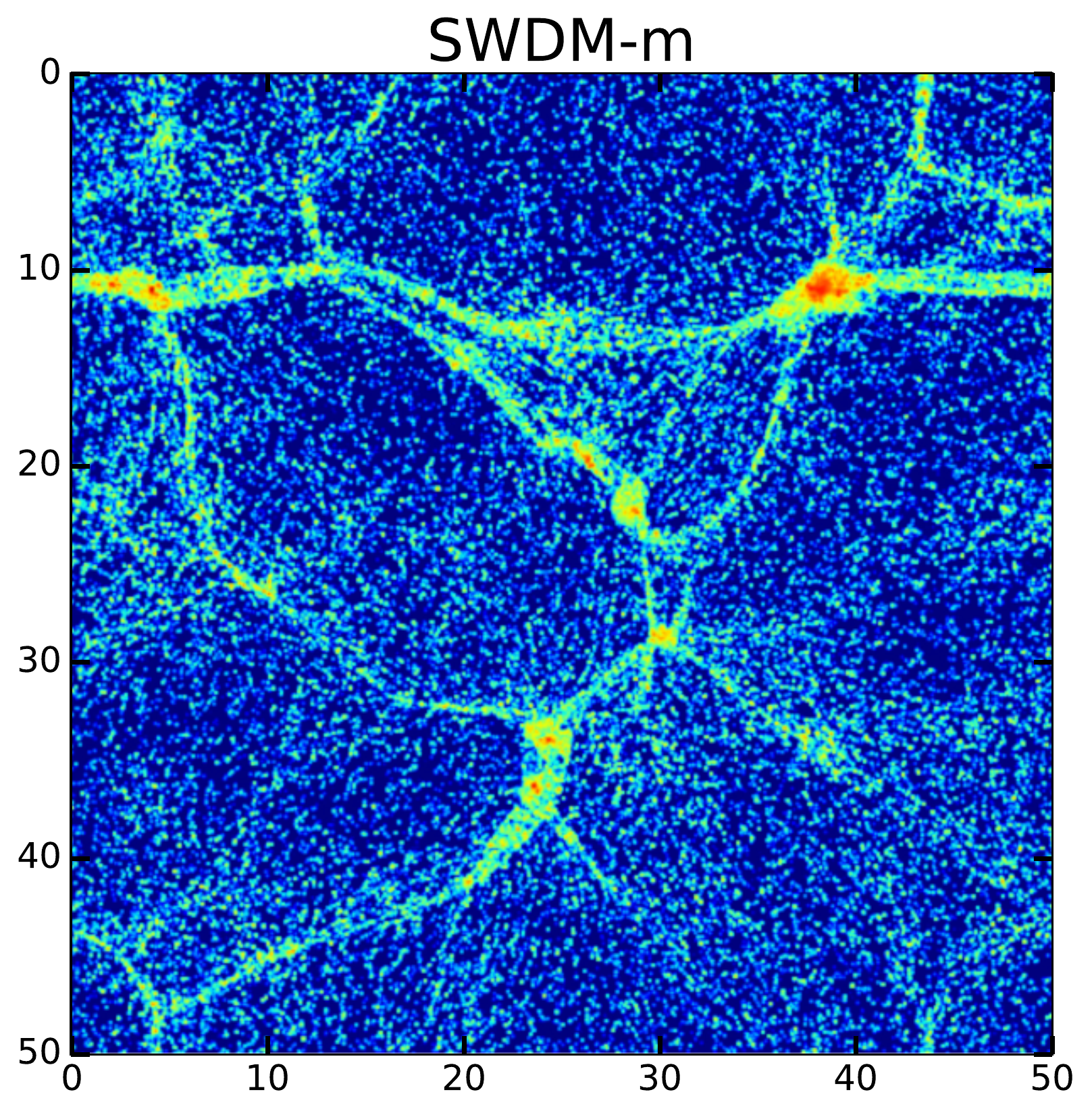}
\includegraphics[width=0.31\textwidth]{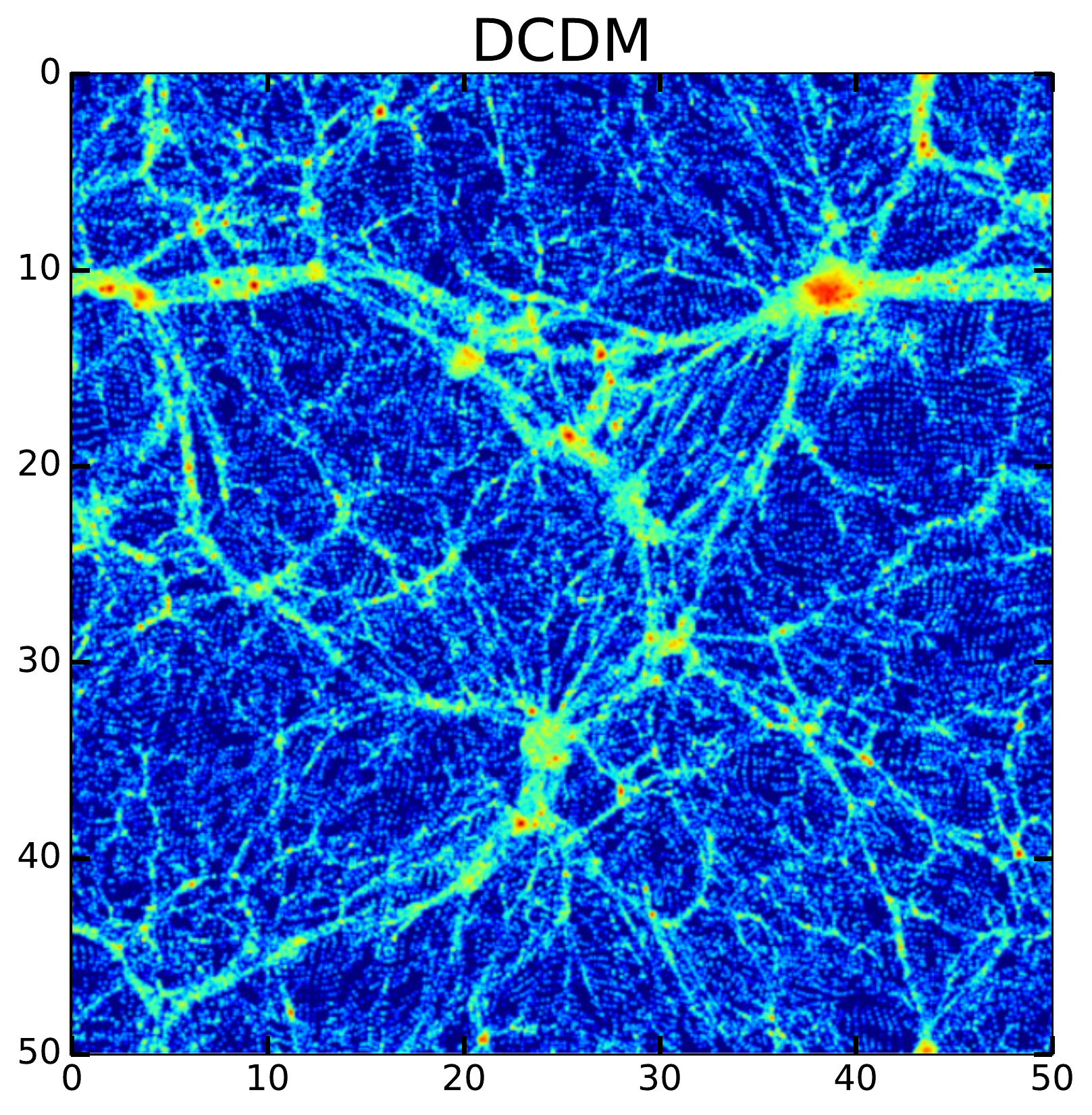}
\includegraphics[width=0.31\textwidth]{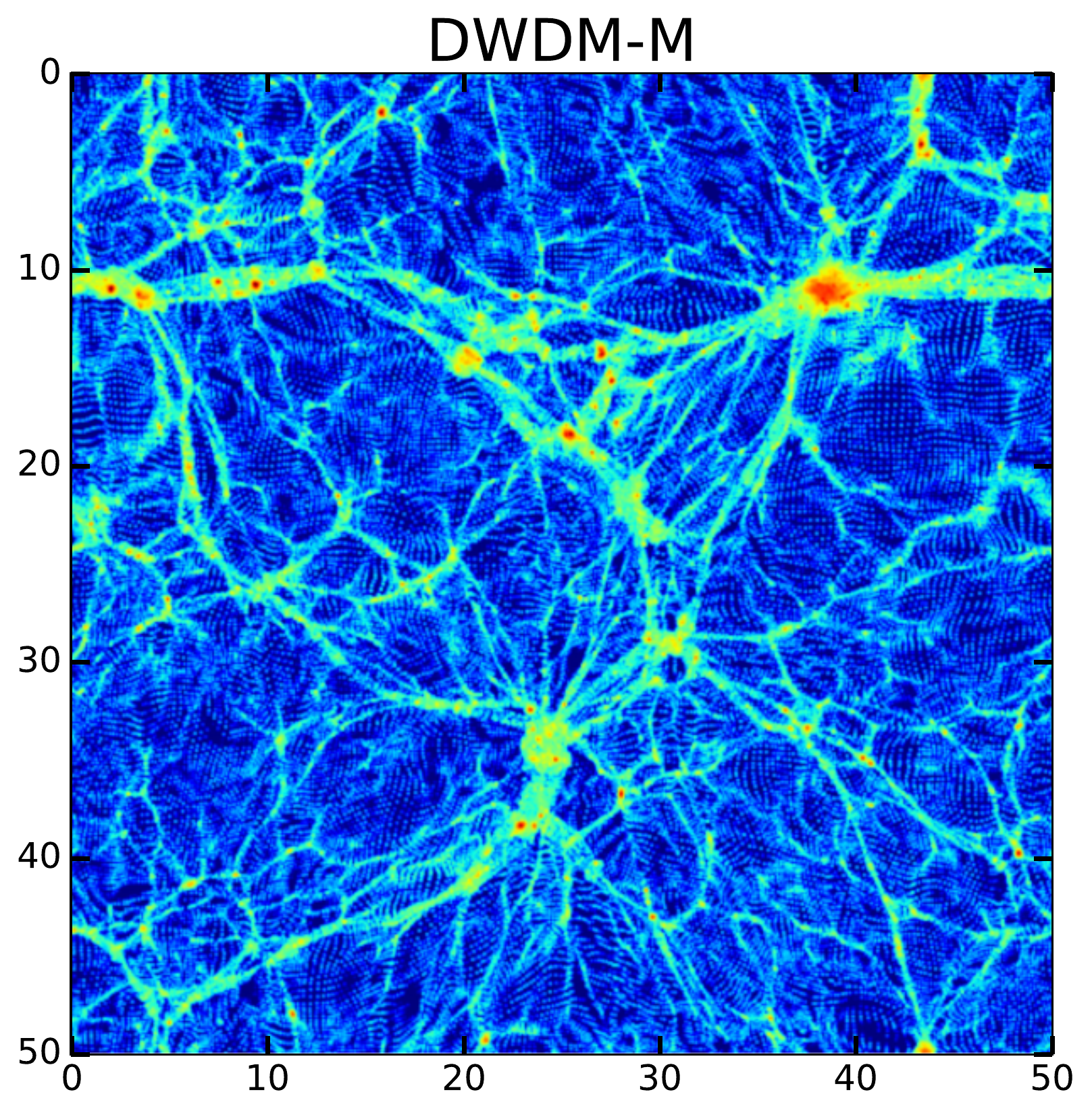}
\includegraphics[width=0.31\textwidth]{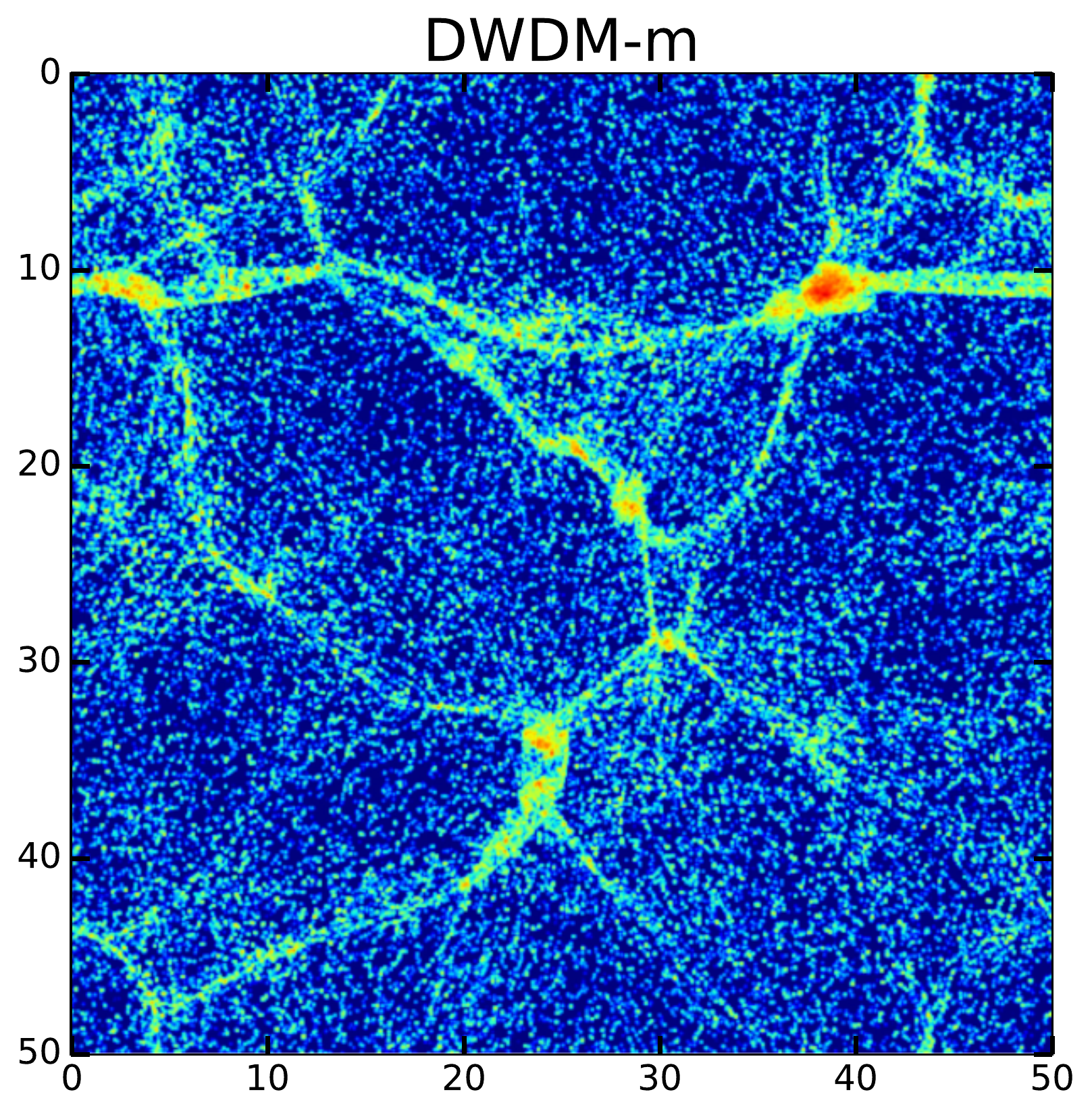}
\includegraphics[width=0.50\textwidth]{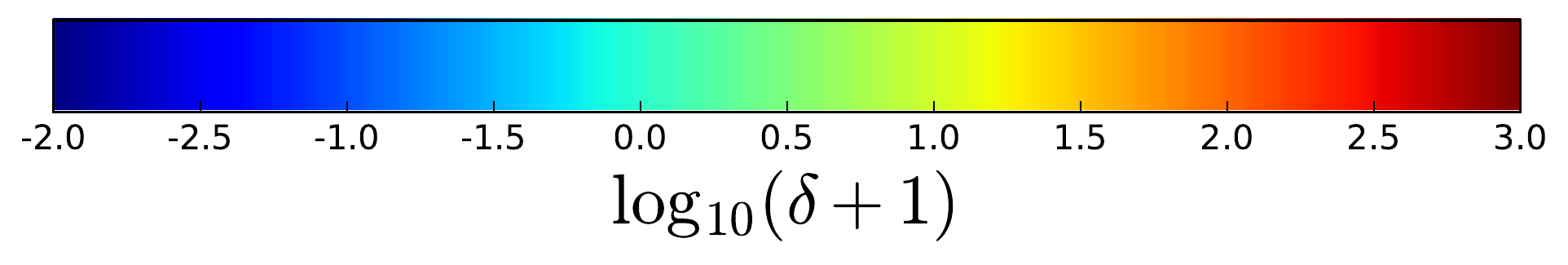}
\caption{Comparison of the density fields at $z=0$. The first and
  second rows correspond to stable and decaying cases (with lifetime
  fixed at the CMB limit, $50\,\mathrm{Gyr}$~\cite{Lattanzi:2007ux}),
  respectively. The first, second and third columns correspond to
  three different paradigms: $\Lambda$CDM, and WDM with masses
  $m_J = 1.5\,\mathrm{keV}$ and $m_J = 0.158\,\mathrm{keV}$,
  from left to right.  The horizontal and the vertical axis are given
  in units of $h^{-1}\, \Mpc$ and represent the size of the
  simulation box.  One clearly sees the free-streaming effect of WDM,
  indicated by the suppression of structure in the density field of
  the WDM simulations.}
\label{Fig:densityfield}
\end{figure}

\begin{figure}
\centering
\includegraphics[width=0.31\textwidth]{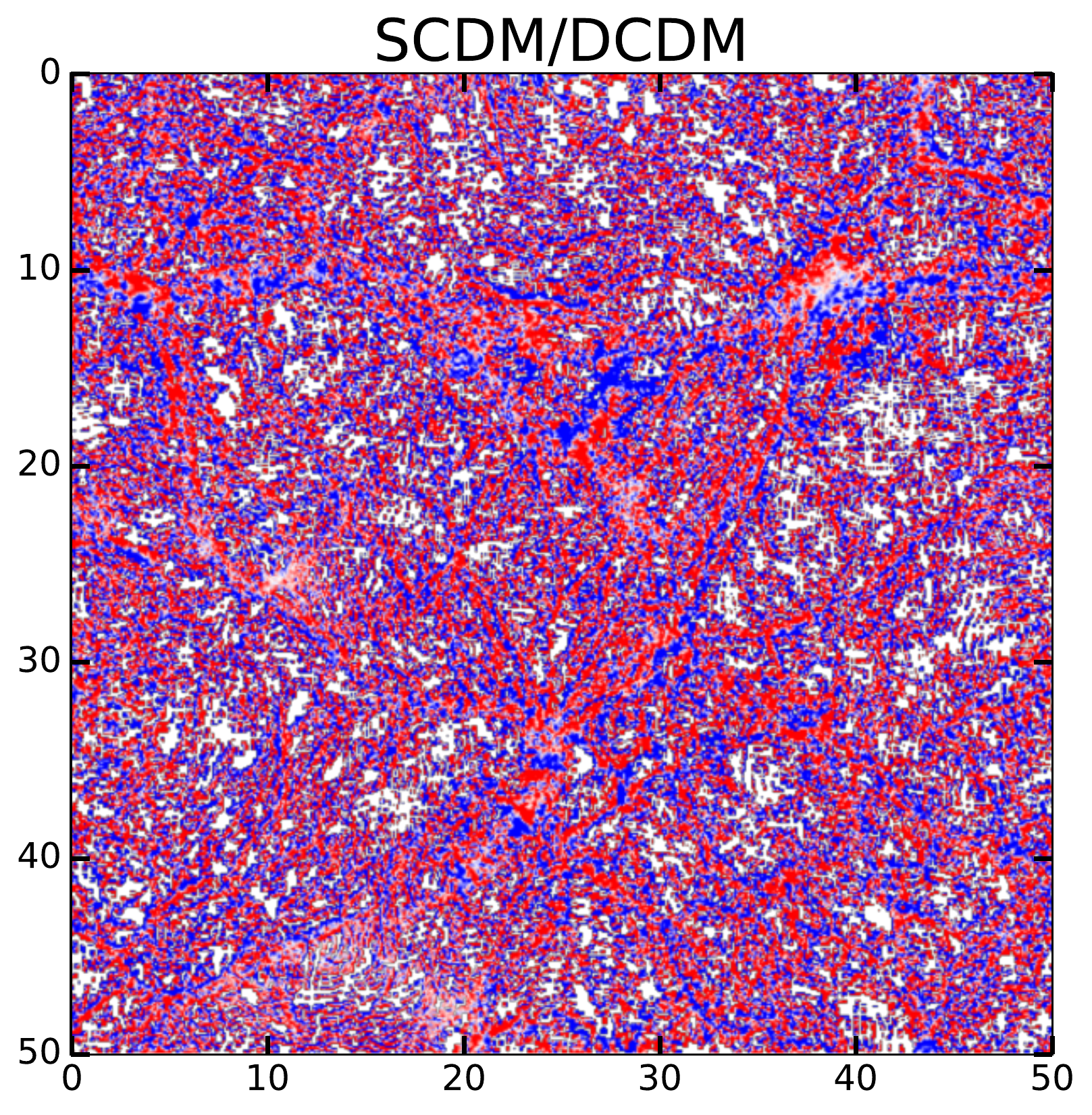}
\includegraphics[width=0.31\textwidth]{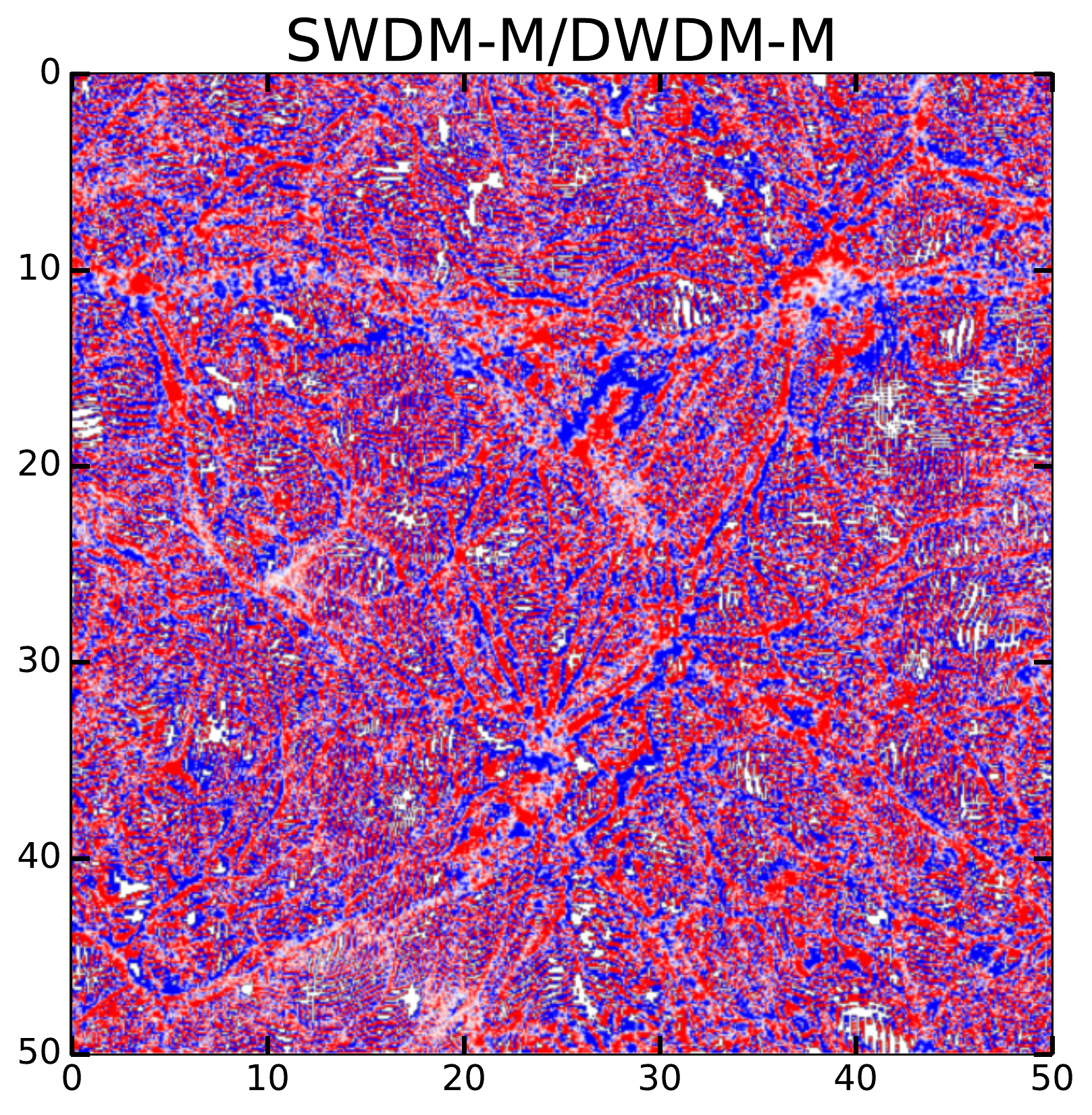}
\includegraphics[width=0.31\textwidth]{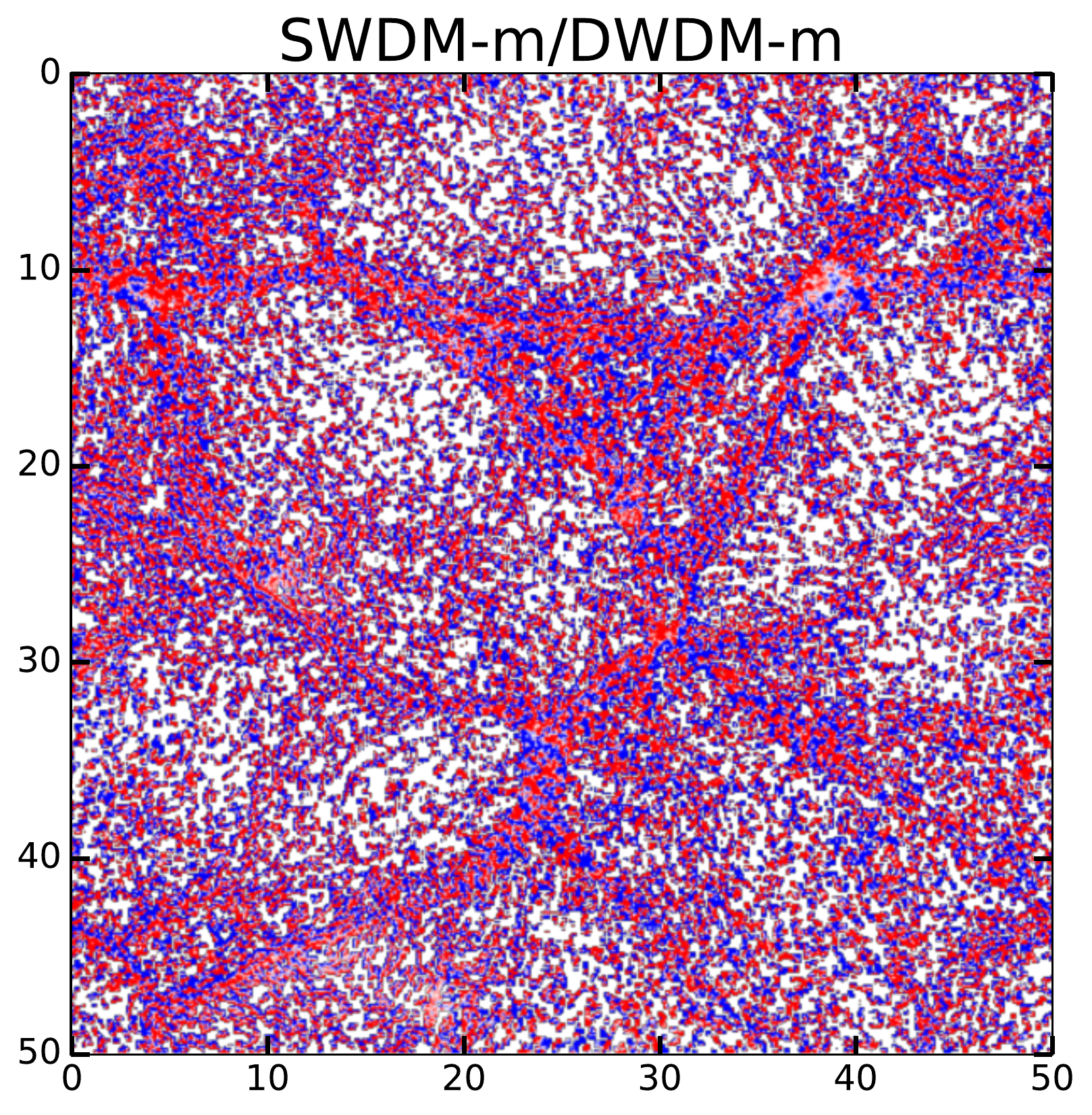}
\includegraphics[width=0.50\textwidth]{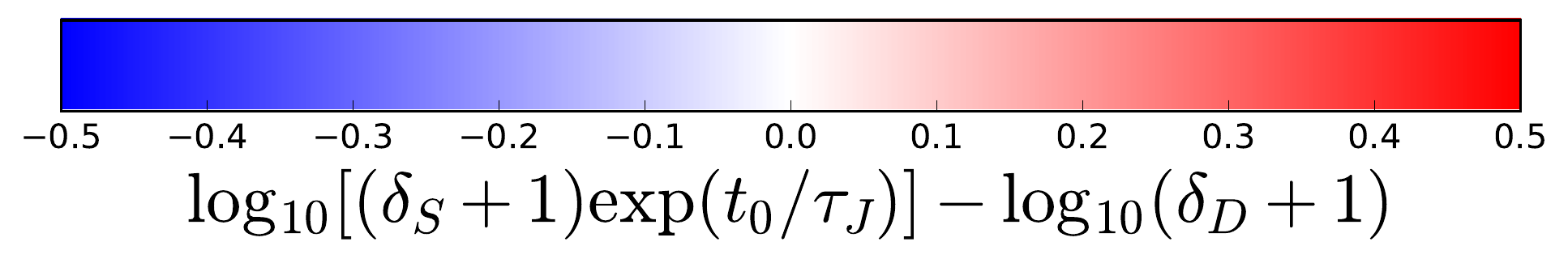}
\caption{Comparison of the relative density fields at $z=0$. The
  left figure represents the relative density field of SCDM and DCDM,
  the middle figure of is the relative density field of SWDM-M and
  DWDM-M, and the right figure is the relative density field of SWDM-m
  and DWDM-m, respectively. The horizontal and vertical axis are given
  in the same units as in Fig.~\ref{Fig:densityfield}. One sees that
  in most of the regions, the density is larger in the stable
  case. However, there are small changes due to subtle features of
  the decay scenario, see text for explanation.}
\label{Fig:rel_densityfield}
\end{figure}

In Fig.~\ref{Fig:densityfield}, we compare the density field extracted
from different simulations. The first, second and third columns
  correspond to three different scenarios: CDM, WDM with mass
  $m_J = 1.5\,\mathrm{keV}$ and WDM with mass
  $m_J = 0.158\,\mathrm{keV}$.  The density field is calculated
from the particle distribution by using the triangular shaped cloud
scheme and further smoothed by a Gaussian filter.  The first and
second rows in Fig.~\ref{Fig:densityfield} correspond to the stable
 and the decaying case.  The density contrast $\delta$ is defined
as
\begin{equation}
\label{density_contrast}
\delta = \dfrac{\rho}{\bar{\rho}} -1,
\end{equation}
where $\rho$ is the local density and $\bar{\rho}$ is the average
density.  The color scale we use in Fig.~\ref{Fig:densityfield} is the
logarithm of $\delta+1$, which represents the ratio of local density
$\rho$ to the average density $\bar{\rho}$. With the density field and
the color scale, one can see how different cosmic structures form in
different cosmologies.  By comparing the stable $\Lambda$CDM and the
stable WDM simulations, one can clearly see the suppression of
structure in the SWDM case, due to the associated free-streaming
effect.  However, the effect of decay is not obvious through
a simple visual comparison of the corresponding stable and decaying density fields.

The well-known suppression of small-scale structure characteristic of WDM is evident when comparing
the different columns in Fig.~\ref{Fig:densityfield}.
We can see that a large portion of small-scale structure is smoothed out in the WDM simulation with WDM mass $m_J = 0.158\,\mathrm{keV}$, due to the large free-streaming length. In fact, the free streaming wavenumber is only one order of magnitude larger than the fundamental mode of the box. On the other hand, the density fields of the $\Lambda$CDM simulations and those of the WDM simulations with WDM mass $m_J = 1.5\,\mathrm{keV}$ look quite similar, because on the scales probed by the simulations
free streaming is rather weak for such cases. However, a lack of small-scale power in the WDM simulation can still be observed.
   
Note that the small peaks in the density field of WDM simulations are due to spurious halos from finite resolution effects and numerical fragmentation, as discussed in Sec.~\ref{Subsec:convergence}. We will also discuss such effects on the halo mass function in Sec.~\ref{Subsec:HMF}.

It is difficult to appreciate the effect of decay by
  performing a quick visual comparison of the density fields in Fig.~\ref{Fig:densityfield}.
Thus, in order to better isolate the effect of decay, we
  refer to the {\sl relative density field} $\rho_S/\rho_D$ of the stable (S) over the decaying (D) case, shown in Fig.~\ref{Fig:rel_densityfield}.
In that figure, the color scale refers to $\log_{10}{\rho_S/\rho_D} = \log_{10}[(\delta_S+1)\exp(t_0/\tau_J)] -  \log_{10}(\delta_D+1) $. 
One can see that the decay
effect reduces the density in most regions of the density field, especially near 
the center of halos and the interior of filaments.
This follows from the change in the gravitational potential due to the
decay, which makes the potential wells more shallow.
This is reflected in the fact that most regions are (relatively) overdense in the stable case, as indicated by the reddish regions in Fig.~\ref{Fig:rel_densityfield}.
Note also that changes in the gravitational potential also affect the dynamics of the
simulation, causing diffusion of the simulation particles. This makes the final density distribution more diffuse with respect to the stable case.
Therefore, we can see that regions near the periphery of
the halos and filaments are denser in the decaying DM case, and appear as the blu-ish regions in Fig.~\ref{Fig:rel_densityfield}.

\subsection{Matter Power Spectrum}

\begin{figure}[h]
\centering
\includegraphics[width=0.45\textwidth]{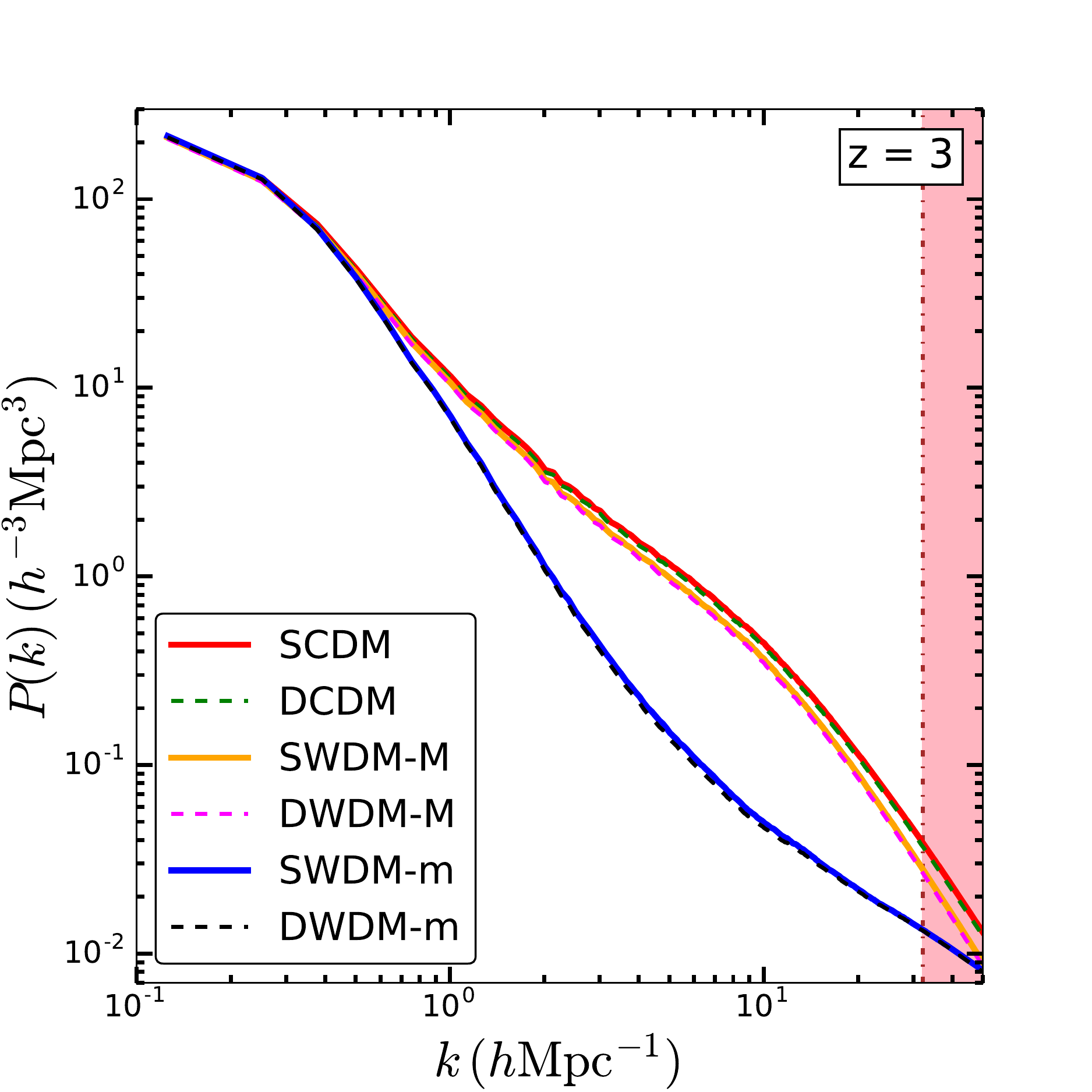}
\includegraphics[width=0.45\textwidth]{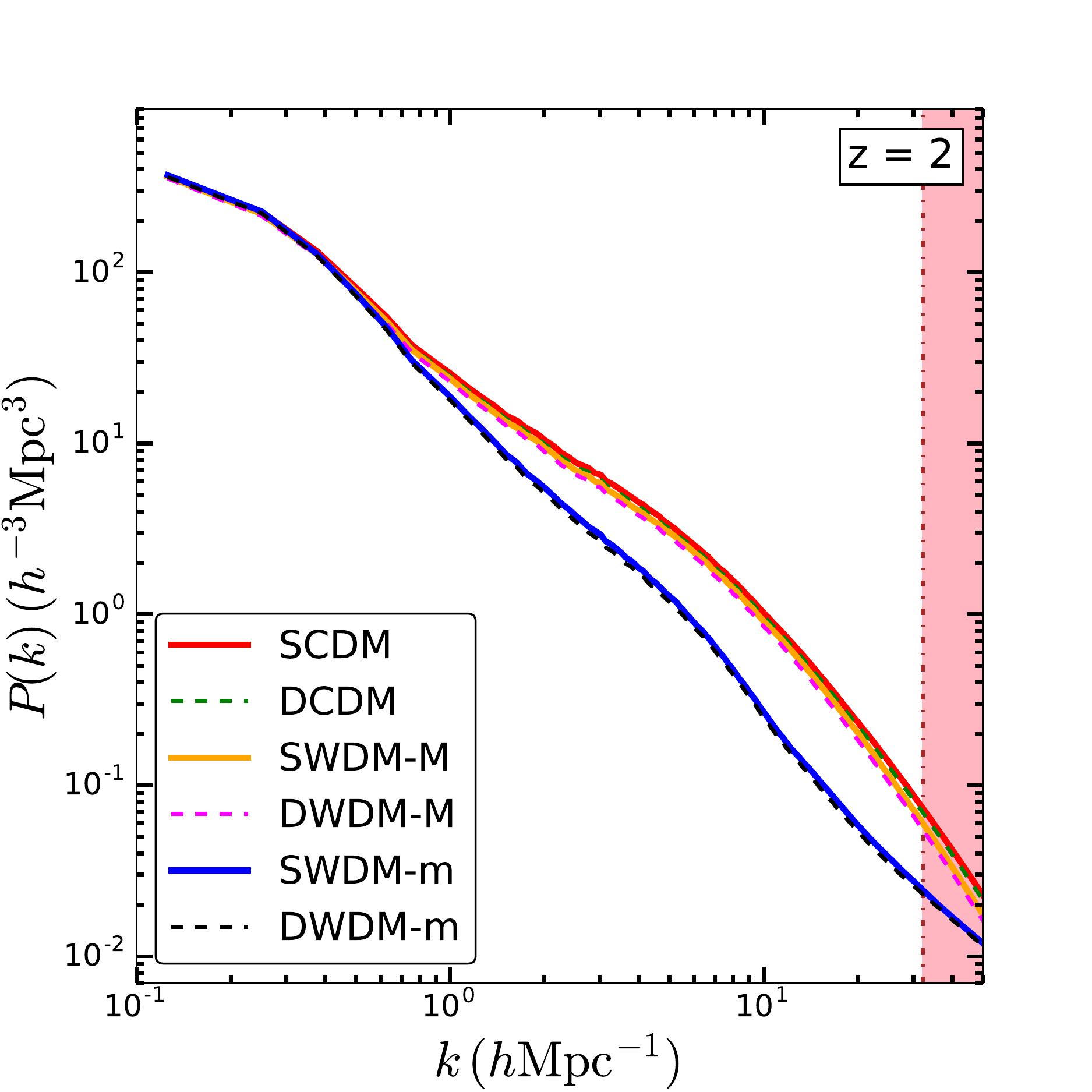}
\includegraphics[width=0.45\textwidth]{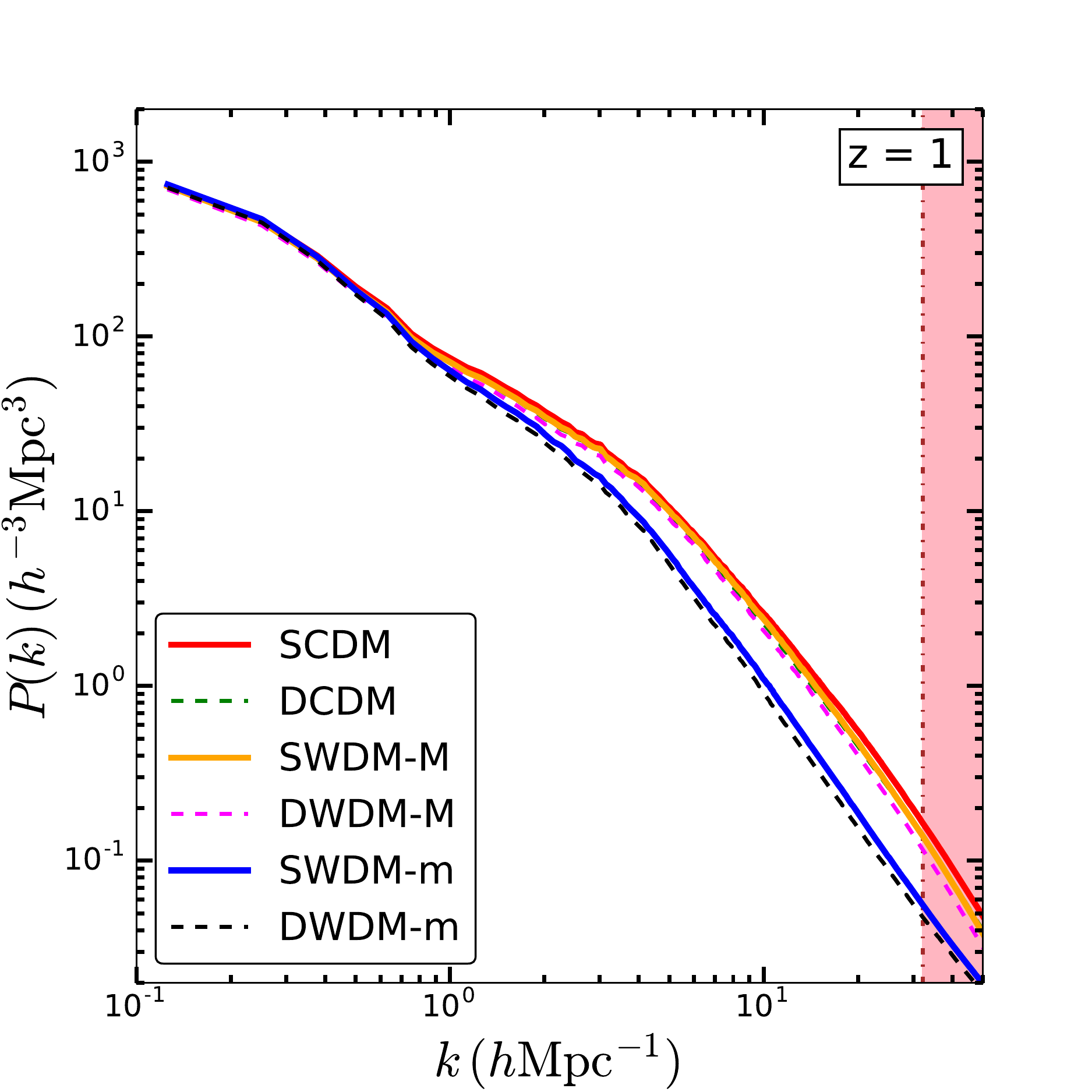}
\includegraphics[width=0.45\textwidth]{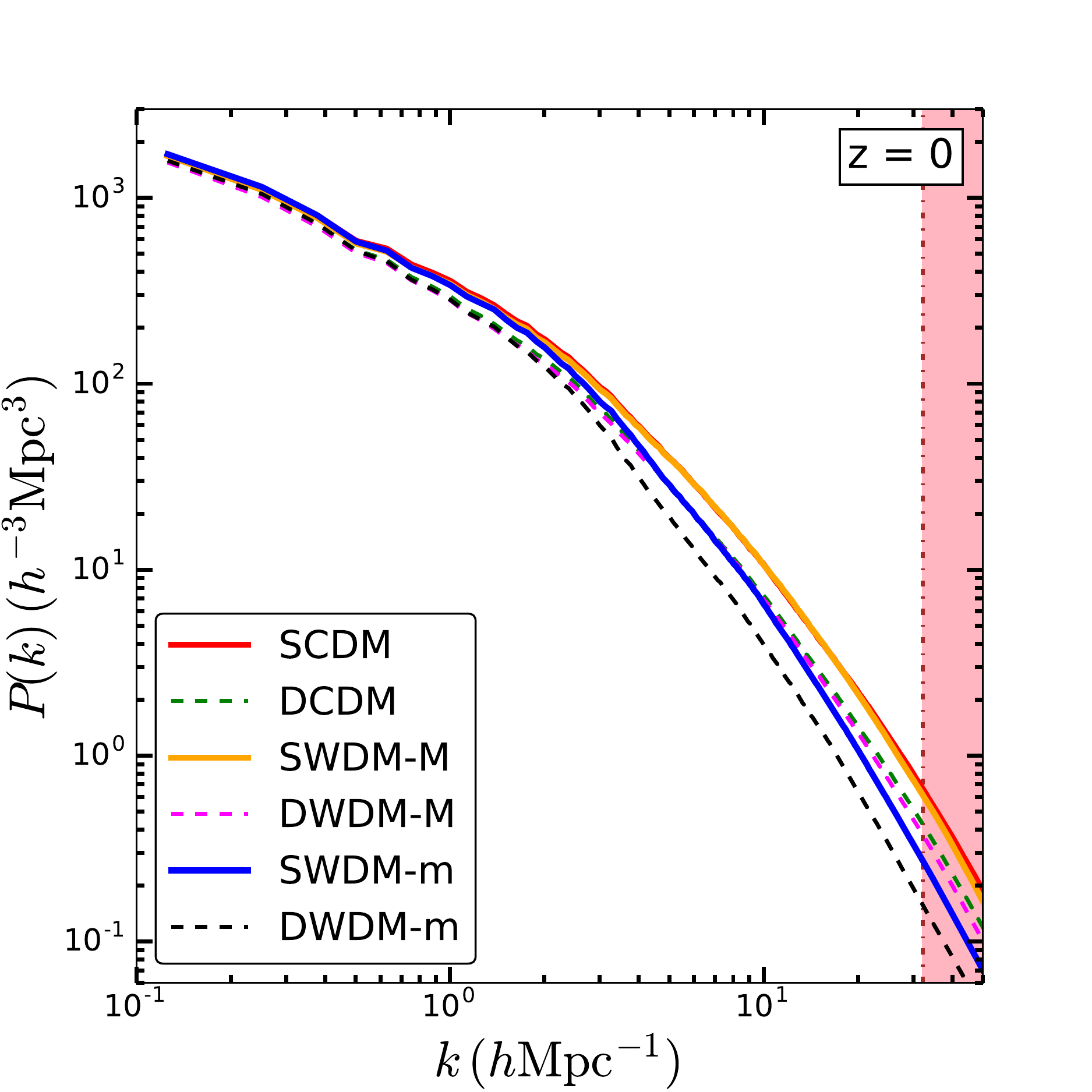}

\caption{Matter power spectra derived from our simulations, for the standard $\Lambda$CDM$\equiv$SCDM, DCDM, SWDM-M, DWDM-M, SWDM-m and DWDM-m cases,
  at redshifts $z=0, 1, 2, 3$. The solid lines represent the stable
  case, while the dashed ones correspond to the decaying case. The
  different colors are associated to different DM mass, and the pink
  band represents length scales smaller than the Nyquist limit.  One can
  clearly see the evolution of the matter power spectrum, as well as
  late-time decay effects.  Further details are given in the text.  }
\label{Fig:Pkcompare}
\end{figure}

\begin{figure}[b]
\centering
\includegraphics[width=0.45\textwidth]{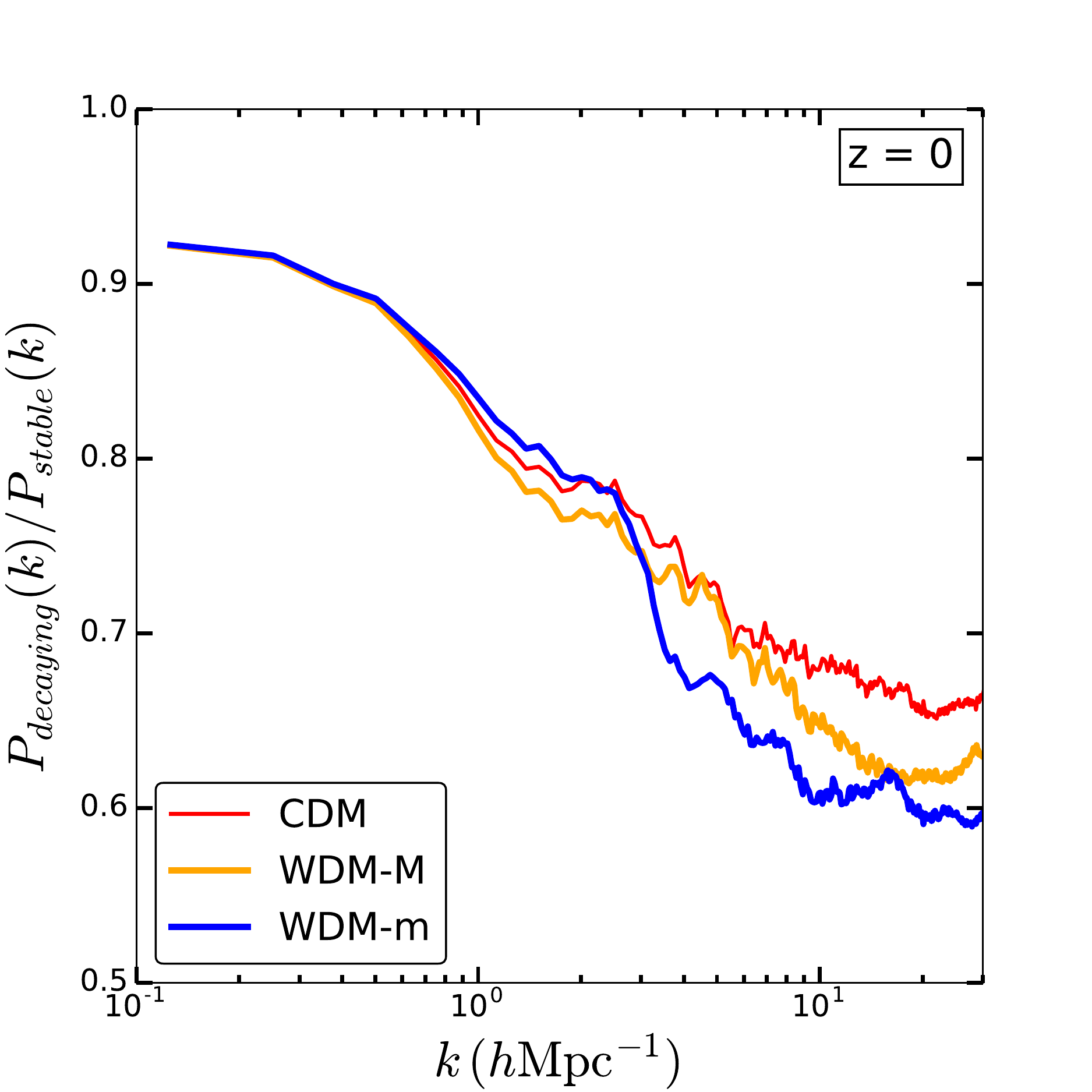}
\includegraphics[width=0.45\textwidth]{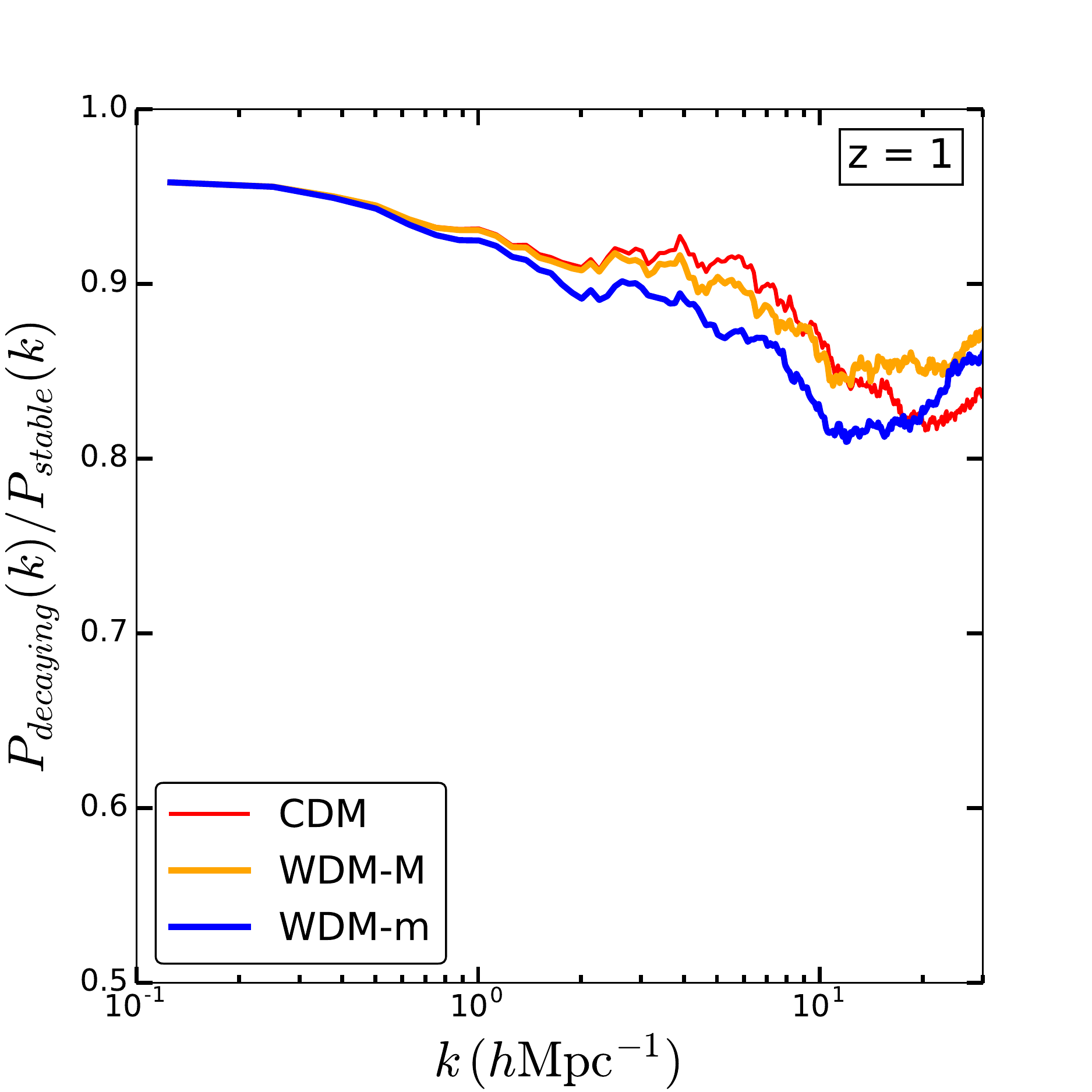}
\caption{Left panel: Ratio between the matter power spectra of decaying and stable dark matter, for CDM (red), WDM-M (orange), WDM-m (blue) at $z=0$. Right panel: Same as the left panel, at $z=1$.
It can be noticed how the effect of the decay is manifest on all scales,
but is more evident on small scales, and also more evident for WDM with respect to CDM.}
\label{Fig:Pk_ratiocompare}
\end{figure}

The matter power spectrum of the simulations is calculated using
the \texttt{ComputePk} code~\cite{2014ascl.soft03015L} with the
triangular shaped cloud scheme.
Note that, since in the simulation we neglect the decay-produced
neutrinos, we only consider the overdensity of the
DM and baryons in calculating the matter power spectrum. In
Fig.~\ref{Fig:Pkcompare}, we show the matter power spectrum
at $z=\{0,\,1,\,2,\,3\}$
for each of the simulations that we have performed,
focusing on the differences between stable and decaying DM
cases.
The solid lines represent the matter power spectra from the
simulations with stable DM, while the dashed lines correspond to the
matter power spectra obtained in the simulations with decaying DM.  
The dashed vertical line corresponds to the Nyquist wavenumber $k_\mathrm{Nyq}$ defined in Eq.~(\ref{eq:nyq}), i.e.
the scale of the average interparticle distance -- the
resolution limit of our simulations.  For our simulation parameters, $k_\mathrm{Nyq}\simeq 32\,h\Mpc^{-1}$.

From Fig.~\ref{Fig:Pkcompare}, one can easily see that the effect of decay becomes manifest at lower redshifts.
This is due to the late decay time of the DM
candidate\footnote{ In our DWDM picture the late majoron decays
  simply reflect the tiny neutrino mass~\cite{Schechter:1981cv}.}.
To quantify the overall effect of decay, we focus on the matter power
spectrum at $z=0$. As a reference, the scale of non-linearity at $z=0$ is roughly $0.15\,h\Mpc^{-1}$.
By comparing SWDM-m and SWDM-M with the standard
$\Lambda$CDM$\equiv$SCDM paradigm, one can see that the matter power
spectra on large scales (small $k$) are identical, but differ on small scales, 
the SWDM spectra being suppressed due to the free-streaming effect of WDM.
This effect is very evident for the SWDM-m case, that has the larger free-streaming length.
The difference between SWDM-M and SCDM  is instead small, and visibile only at
the largest k's, because
the free-streaming length for WDM with $m_J = 1.5\,\mathrm{keV}$ is still quite small,
and thus free-streaming does not cause too much suppression on the scales probed in our simulations.
By comparing DWDM-m and DWDM-M with SCDM, one can see that the small
scale suppression due to the free-streaming effect of WDM still
exists.  Moreover, there is further suppression on all scales caused
by the effect of decay. The presence of the decay, inherent in the BV
model~\cite{Berezinsky:1993fm}, reduces the matter energy density in
the universe, hence the growth factor is reduced, which delays the
formation of structure. The decay-induced suppression
does not show a strong dependence on scale. 
This should contrasted with the free-streaming effect of WDM, that has a strong dependence
on the scale, due to the scale of the cut-off in the initial transfer
function, related to the mass and temperature of the WDM.
A lighter thermal WDM will cause a cut-off in the matter power spectrum on a
larger scale, as originally envisaged in the BV model.  

The suppression due to the decay seems however to gradually decrease towards large scale.
To better assess this behaviour, we compute the ratio between the decaying and stable power spectra, 
$P_{decaying}(k)/P_{stable}(k)$, for each of the three pairs of simulations. These are shown,
for $z=0$ and $z=1$,
in Fig.~\ref{Fig:Pk_ratiocompare}, that can be compared with Fig.~2 in Ref.~\cite{Enqvist:2015ara},
that shows the same quantity for the CDM case. First of all, we note that our results for CDM are very consistent with those of 
Ref.~\cite{Enqvist:2015ara}, also considered the slighthly different values for the lifetimes between their work and the present one.
Let us briefly discuss the features of the curves in  Fig.~\ref{Fig:Pk_ratiocompare}. 
We find first of all confirmation that the decay suppresses 
the spectrum on all scales under consideration, since the curves always lie below unity.
On the largest scales shown in the plot, the three ratios converge to a constant common value. That the large-scale behaviour of the curves is common could be easily expected by noting that, above the free-streaming length, WDM and CDM behave the same way. 
On the other hand, the suppression due to the decay is more evident on small scales, in all the cases under consideration.
As noted by Enqvist et al.~\cite{Enqvist:2015ara}, this is due to the fact that mode-mode couplings make 
that differences that are small in the linear regime get enhanced by the nonlinear evolution. In fact, we can see that
this effect is less pronounced in the right panel of Fig.~\ref{Fig:Pk_ratiocompare}, corresponding to $z=1$, 
when more scales were still in the linear regime with respect to $z=0$ (shown in the left panel).

Another interesting feature that can be noticed in Fig.~\ref{Fig:Pk_ratiocompare} 
is that the nonlinear enhancement of the effect of the decay on small scales 
is stronger for lighter WDM. There is a distinct drop in the curve, 
especially evident in the $m_J=0.158\,\mathrm{keV}$ case, at scales right above the free-streaming length.
In other words, it seems that the combination of the cutoff in the linear power spectrum due to the WDM
thermal velocity and of the nonlinear evolution, enhances the effect of the DM decay.

To summarize, by comparing the decaying and the stable cases, one can
see that the effect of decay is to suppress power on all scales,
with the suppression being more severe on the small, nonlinear scales.
In contrast, the effect of WDM {\sl per se} is to suppress the matter
power spectrum on small scales, depending on the mass of the WDM
candidate. Also, the small-scale suppression due to the decay is more evident for WDM, 
as one can see by comparing the matter power spectrum of
$\Lambda$CDM simulations with those corresponding to WDM simulations.

\subsection{Halo Mass Function}
\label{Subsec:HMF}

\begin{figure}
\centering
\includegraphics[width=0.45\textwidth]{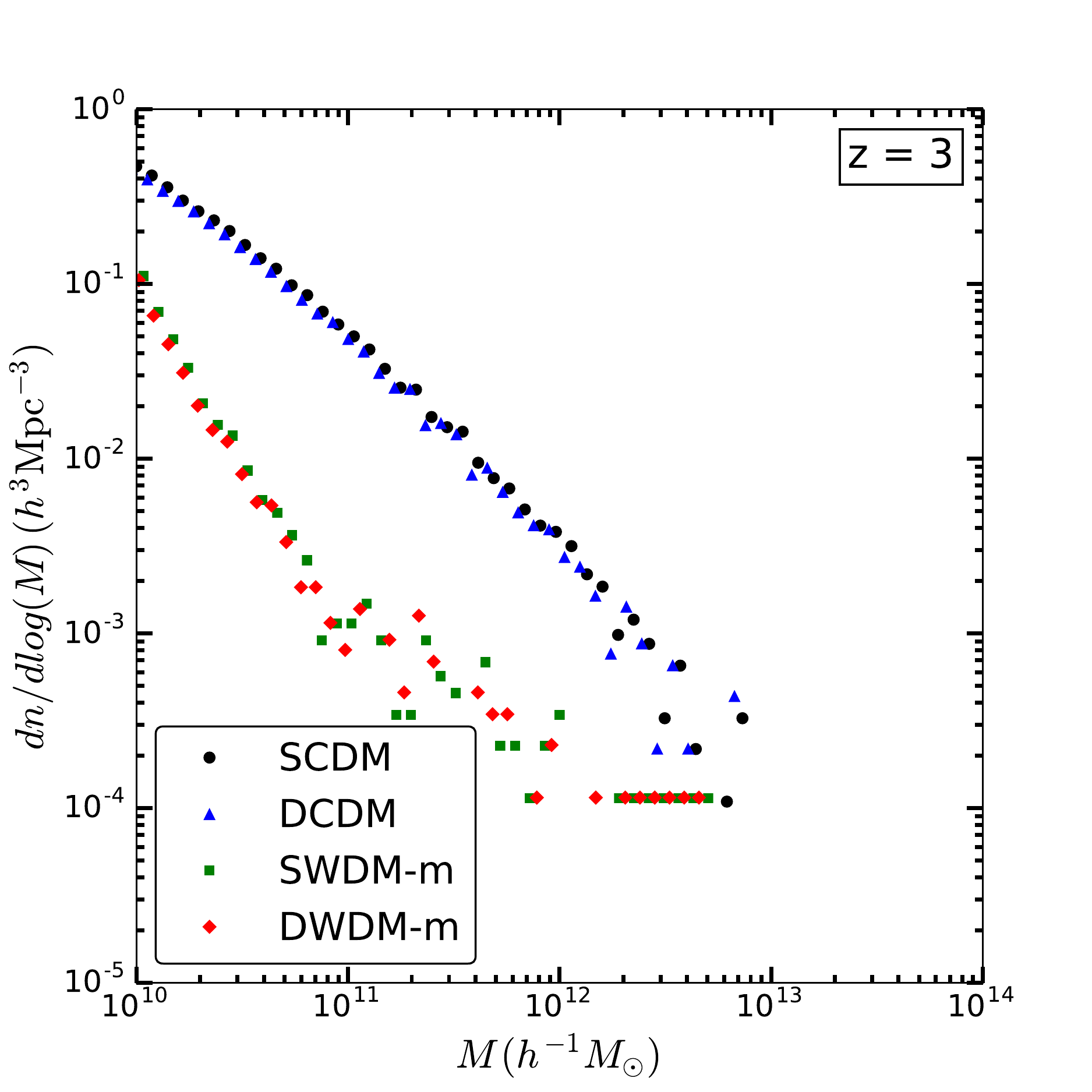}
\includegraphics[width=0.45\textwidth]{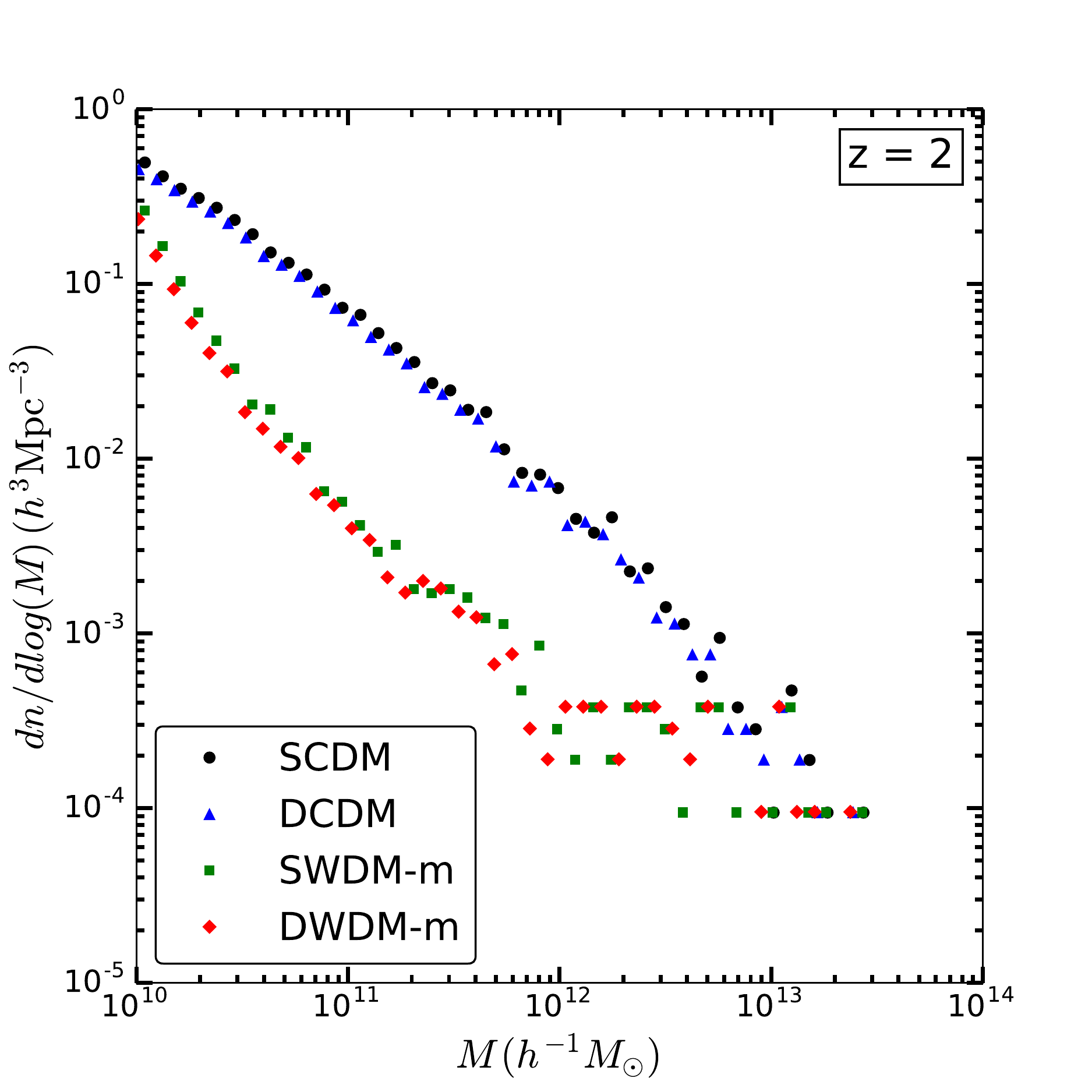}
\includegraphics[width=0.45\textwidth]{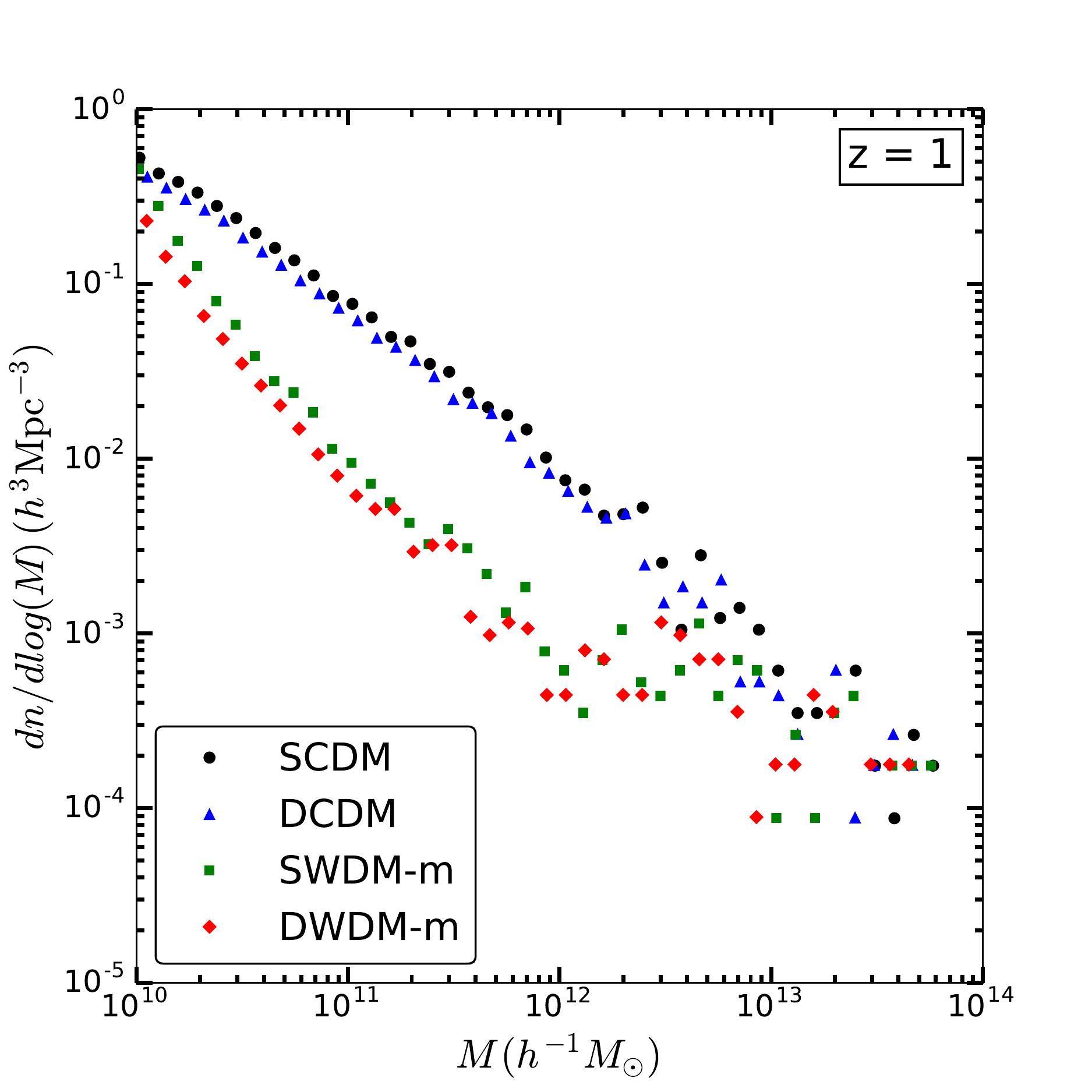}
\includegraphics[width=0.45\textwidth]{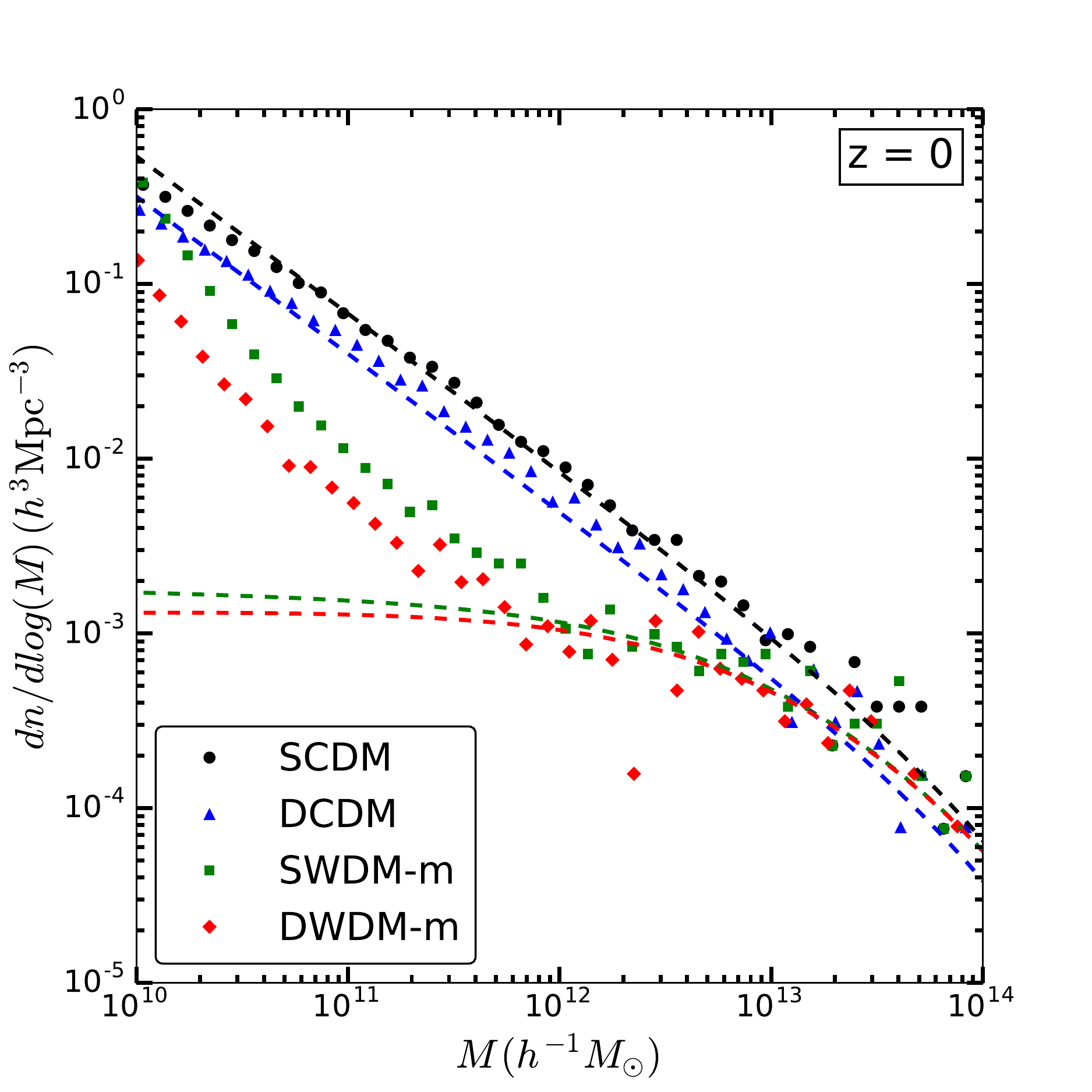}
\caption{  Evolution of the halo mass function for the standard
  $\Lambda$CDM$\equiv$SCDM paradigm (black circle) compared with the
  simulations corresponding to DCDM (blue triangle), SWDM-m 
  (green square), and DWDM-m (red diamond).
 The dashed lines of corresponding colors in the $z=0$ panel 
represent our derived halo mass function fits based on the 
given cosmology and the data points obtained from our simulations.
The data points of our simulations that do not fit well to
  the theoretical WDM halo mass function are mainly due to spurious halos.
By comparing the stable and decaying cases, we can see that the 
effect of decay is to reduce the number density of halos 
for all mass scales.
However, the effect of the warm DM nature is seen by setting a cut-off
mass, which is the mass scale that the halo mass functions
of WDM simulations start to deviate from those of CDM simulations. }
\label{Fig:Halocompare}
\end{figure}

The halo mass function is defined as the number density of DM halos
per unit logarithmic mass interval. In order to estimate the halo mass function,
we need to identify halos, i.e., bound objects, within the large set of particles in our simulations.
For this purpose, we make use of the parallel halo finder package $\texttt{AHF}$~\cite{Knollmann:2009pb} in
order to calculate the halo mass function based on our snapshots of
simulations.
In $\texttt{AHF}$, the adaptive mesh refinement algorithm is adopted
to identify clumps in the density field.
Therefore, it can build up the hierarchical structure for the halos
and sub-halos obtained in the snapshots. 
After iteratively removing the particles unbounded by the
gravitational potential of the halo and refining the halo edge, the
properties of the  halos are finally determined by the particles within
its virial radius $R_{vir}$.
Here $R_{vir}$ is defined as the point where the density profile of
the particles drops below $\Delta_{vir} \rho_c$ in which
$\Delta_{vir}$ is a constant depending on the cosmology and the
$\rho_c$ is the critical density of the universe.
In Fig.~\ref{Fig:Halocompare}, the black circle, blue triangle, green
square, and red diamond represent halo mass functions obtained from
the $\Lambda$CDM$\equiv$SCDM paradigm and the simulations
corresponding to DCDM, SWDM-m and DWDM-m, respectively.
In order to avoid cluttering in the figures, we do not illustrate
the halo mass functions corresponding to DWDM-M and SWDM-M in the figure.
The dashed lines of corresponding colors in the $z=0$ panel are
  the halo mass function fits calculated based on the given cosmology
  and the halo mass functions obtained from our simulations.

Note that as a result of the strong cut-off in the WDM transfer function, resulting in a suppression of small-scale power, discreteness effects close to the resolution limit will be more important than for CDM simulations.
Indeed, as a consequence of finite resolution effects in the simulations,  
spurious clumps are produced by numerical fragmentation~\cite{Wang:2007he,Lovell:2013ola,Leo:2017zff}.
Note that the halos produced in early simulations~\cite{Bode:2000gq} were considered to be the result of the "top-down" structure formation scenario of WDM 
~\cite{Knebe:2003hs}. 
However, further studies have demonstrated that this phenomenon depends on the average interparticle distance, i.e. on the resolution of the simulation, hence could be regarded as a numerical artifact~\cite{Wang:2007he}.
For small halo masses, these spurious clumps will outnumber the genuine halos.

In order to identify the spurious clumps, we first calculate the halo
mass function of the corresponding cosmology using the code
\texttt{hmf},
which is the back end of \texttt{HMFcalc}~\cite{Murray:2013qza}, with
the fitting model in Ref.~\cite{Tinker:2008ff}.
Then we introduce the fitting method in Ref.~\cite{Schneider:2011yu},
which provides a precise fit for the halo mass function of WDM
cosmology.
The overall halo mass function fit for the WDM scenarios used in this
work can be written as
\begin{equation}
n(M)= (1+M_{hm}/M)^{-\gamma}\times n_{Tinker}(M),
\end{equation}
where the $M_{hm}$ is defined as the mass scale at which the amplitude
of the WDM transfer function is reduced to 1/2,
$\gamma$ is a free parameter for fitting the correct shape of the halo
mass function and $n_{Tinker}(M)$ is the halo mass function fit in
Ref.~\cite{Tinker:2008ff}.
$M_{hm}$ is expected to mainly affect the properties of WDM
haloes~\cite{Schneider:2011yu}.
$M_{hm}$ is related to its corresponding length scale $\lambda_{hm}$
by
\begin{equation}
M_{hm} = \dfrac{4\pi}{3} \bar{\rho} \left(\dfrac{\lambda_{hm}}{2}\right)^3,
\end{equation}
where $\bar{\rho}$ is the average density of the universe.
Here we follow Ref.~\cite{Schneider:2011yu} to calculate $M_{hm}$
based on our simulation results and perform the fitting.
For the best-fit values, we find that $\gamma \approx 0.309$ in SWDM-m
and $\gamma \approx 0.345$ in DWDM-m, in which the larger $\gamma$ for
DWDM-m indicates the existence of further suppression coming from the
decay.
For the halo mass function fit for DCDM, we simply introduce a factor
$A$ to account for the effect of decay, which is written as
$n(M)=A \times n_{Tinker}(M).$

By comparing the halo mass function fits with the halo mass functions
from the simulations,
we can infer that for SWDM-m and DWDM-m the halo mass functions from
the simulations (green square and red diamond) deviating from the
corresponding halo mass function fits are mainly composed of spurious
halos.
The genuine halo mass functions for WDM simulations should exhibit the
same trend as the corresponding halo mass function fits for small
masses.
The effect of decay and the free-streaming effect of WDM can also be
separated due to their distinct impact on the halo mass functions.
From the difference between the stable case and decaying case, we can
infer that the effect of decay in the halo mass function is to reduce the
number density of halos in every mass scale.
 In other words, the effect of decay produces an overall downward
  shift on the halo mass function.
On the other hand, the free-streaming effect of WDM is to set a
cut-off halo mass which is roughly the mass scale that the WDM halo
mass function deviates from CDM halo mass function.
We also note that, at large halo masses, it is difficult to assess the differences between the halo mass functions of different cosmologies, due to the variance caused by the scarcity of halos with large masses.

\black

\subsection{Effects of baryonic physics}
\label{Sec:baryon}

To close our discussion we now comment on the effects of baryonic physics, so far neglected.
Despite the fact that baryons are themselves biased tracers of the DM gravitational potential,
their role in structure formation is distinct from that of DM as a result of their ability to 
cool down through radiative processes.
This makes baryons able to form compact astrophysical objects, such as stars,
resulting in a different distribution compared to that of DM. 
Thanks to the development of N-body simulation techniques, high-resolution 
and large-scale hydrodynamics simulations have now become feasible.
Many studies have shown that baryonic processes can generate
non-trivial effects on astrophysical observables such as the halo density
profile, the matter power spectrum and the halo mass function. However, the precise details
 of baryonic processes remain poorly understood~\cite{Rudd:2007zx,Stanek:2008am,Cui:2011,vanDaalen:2011xb,Bocquet:2015pva,Chan:2015tna,Despali:2016meh}.
Baryons affect the halo mass and density profile in several ways.
When falling into the potential wells created by the DM, baryons are gravitationally heated and exchange energy with DM during
 relaxation. Hence they remain more diffuse and can create core-like density profiles at the center of halos.
Later, as they dissipate energy through radiative processes, baryons 
sink into the center of halos and finally convert into stars. This steepens the 
density profile near the center.
On the other hand, the presence of supernovae (SN) and active galactic nuclei (AGN), 
reduces the effect of radiative cooling and adiabatic contraction, since 
baryons are ejected out from the center of the halos. 
Baryonic physics also affects the matter power spectrum, see for example Ref.~\cite{vanDaalen:2011xb} for a 
comprehensive review. 
At intermediate scales ($k\simeq 0.8 - 5 \,h\Mpc^{-1}$) the power spectrum is suppressed
because the pressure of baryons smoothens the density field. In contrast,
the spectrum rises at small scales because of radiative cooling that allows baryons to cluster at those scales.
Concerning the halo mass function, the number density of low-mass halos increases as a result of cooling and star formation.
However, AGN feedback can reduce the effect of cooling and hence the abundance of low-mass halos.

In summary, we have shown that in a DDM cosmology, there is an overall suppression of
the density fluctuations, due to the decay of the DM particle. This is accompanied, in the case of WDM, by the well-known smoothing of the density field and suppression of the small-scale power in the matter power spectrum and halo mass function.
Taking into account the full set of baryonic processes, including gravitational heating, radiative cooling, 
adiabative contraction, stellar and AGN feedback, we expect that the impact of baryonic 
physics will be generally weakened by the decaying and warm nature of DM.
In fact, since the gravitational potential is shallower in the DWDM scenario, due to both the decay and the free-streaming of WDM, the ability of halos to 
accrete baryons into their center will be somehow limited.
Therefore, we expect the star-formation rate to decrease as a result, leading in turn to a lower
efficiency of the stellar and AGN feedback.
In this sense, we would argue that the baryon component remains in a relatively
smooth density distribution containing fewer structures, while still being a biased tracer of the DM.
This is only a qualitative assessment of the effects of baryonic physics in a DWDM structure formation scenario. Full-fledged hydrodynamic simulations would be necessary to assess the interplay between baryons and DWDM.

\section{Conclusions}
\label{sec:conclusion}

In this paper we have examined the cosmology of warm dark matter, both
for the stable as well as decaying cases, paying special attention to
how it affects structure formation.
We have performed DM-only N-body simulations of the nonlinear evolution and compared the matter power spectrum associated to warm dark
matter masses of $1.5\,\mathrm{keV}$ and $0.158\,\mathrm{keV}$, with
that expected for the stable cold dark matter $\Lambda$CDM paradigm,
taken as our reference model.  
We have scrutinized the effects associated to the warm nature of dark
matter, as well as the fact that it decays. We find that the nonlinear evolution
somehow couples the two effects, in such a way that the effect of the decay
becomes more pronounced below the free-streaming scale of WDM.
All of our considerations are general, though we have been
  strongly motivated by the fact that the DWDM scenario can naturally
  appear in particle physics.
  A nice example is provided by the keV majoron DM scenario suggested
  in the original BV-proposal~\cite{Berezinsky:1993fm}.
The majoron itself emerges as a Nambu-Goldstone boson within a broad
class of particle theories where neutrino mass generation takes place
through the spontaneous breaking of a continuous ungauged lepton
number symmetry.
The majoron picks up a mass from gravitational effects, expected to
explicitly break global symmetries. Hence it must necessarily decay to
neutrinos, with an amplitude proportional to their tiny mass, which
typically gives it cosmologically long
lifetimes~\cite{Schechter:1981cv}.
As a reference value for decaying dark matter lifetime we have taken
the conservative limit following from CMB observations obtained
in~\cite{Lattanzi:2013uza}.
We have modified the standard N-body simulation code so as to include
the effect of decay besides the free-streaming effect.
Through these simulations we have shown that the DWDM
picture suggested in the BV proposal leads to
predictions on small scales that differ substantially from
those of the standard $\Lambda$CDM paradigm. A dedicated analysis,
using better resolution simulations and including baryons, is required to assess whether this could address
the potential drawbacks of the $\Lambda$CDM scenario.
We have also qualitatively discussed the possible impact on the DWDM scenario 
when baryonic physics is taken into account.
Our results illustrate that the observations of large-scale structures
in the Universe can in principle be used in order to constrain the
particle physics model underlying the origin of dark matter.
In our case, our results may be extended in order to constrain the
lifetime and mass of the keV dark matter majoron.

\section*{Acknowlegdments}

This research was supported by the Spanish grants FPA2017-85216-P (AEI/FEDER, UE), SEV-2014-0398 and PROMETEOII/2018/165 (Generalitat Valenciana), by the Italian INFN through the InDark and Gruppo IV fundings, by ASI through the Grant 2016-24-H.0 (COSMOS) and the ASI/INAF Agreement I/072/09/0 for the Planck LFI Activity of Phase E2, and by the Taiwan MoST grants MOST-105-2112-M-007-028-MY3 and MOST-107-2112-M-007-029-MY3.\\[.3cm]

\appendix
\section{Results for 5.3~keV decaying dark matter }

\begin{figure}[b]
\centering
\includegraphics[width=0.45\textwidth]{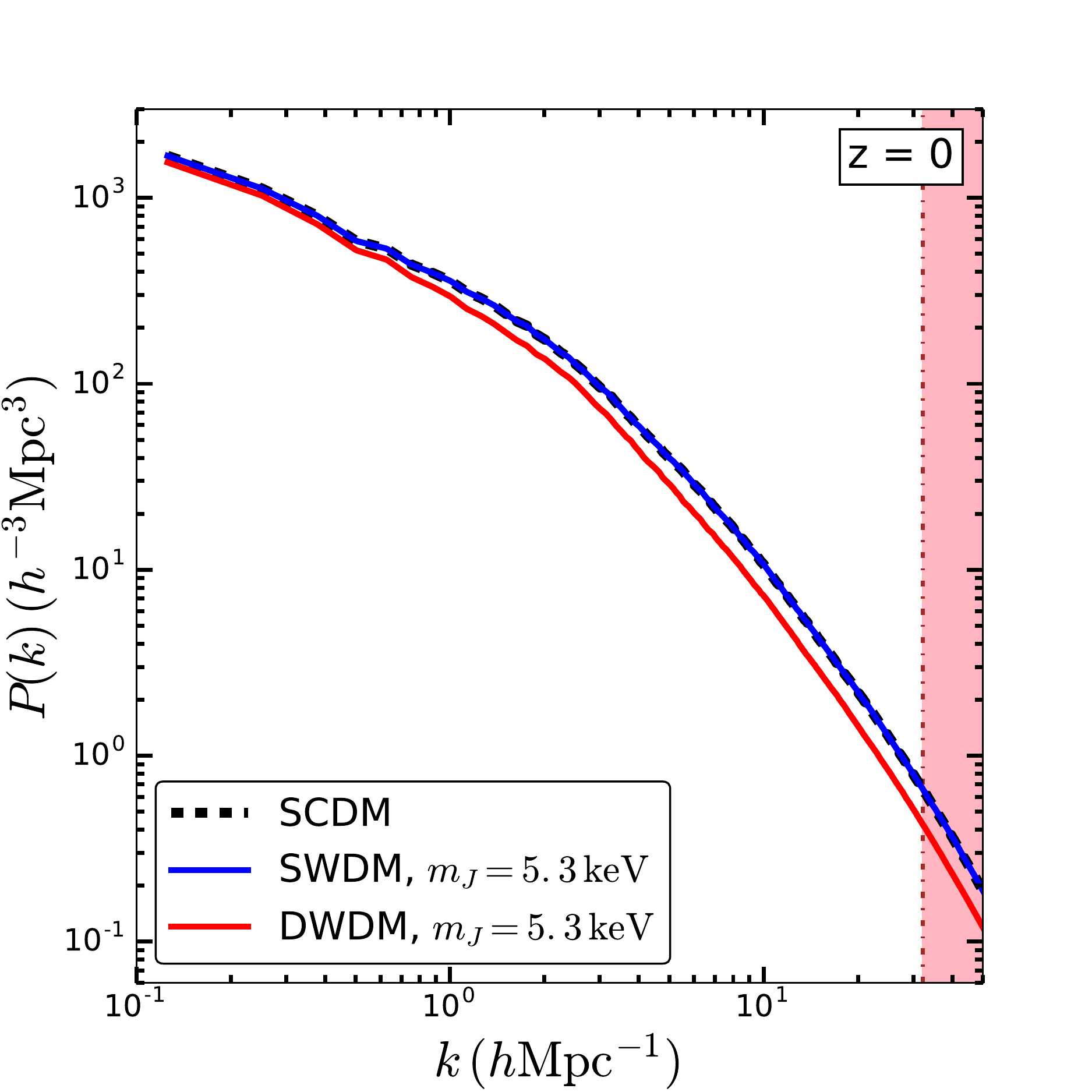}
\includegraphics[width=0.45\textwidth]{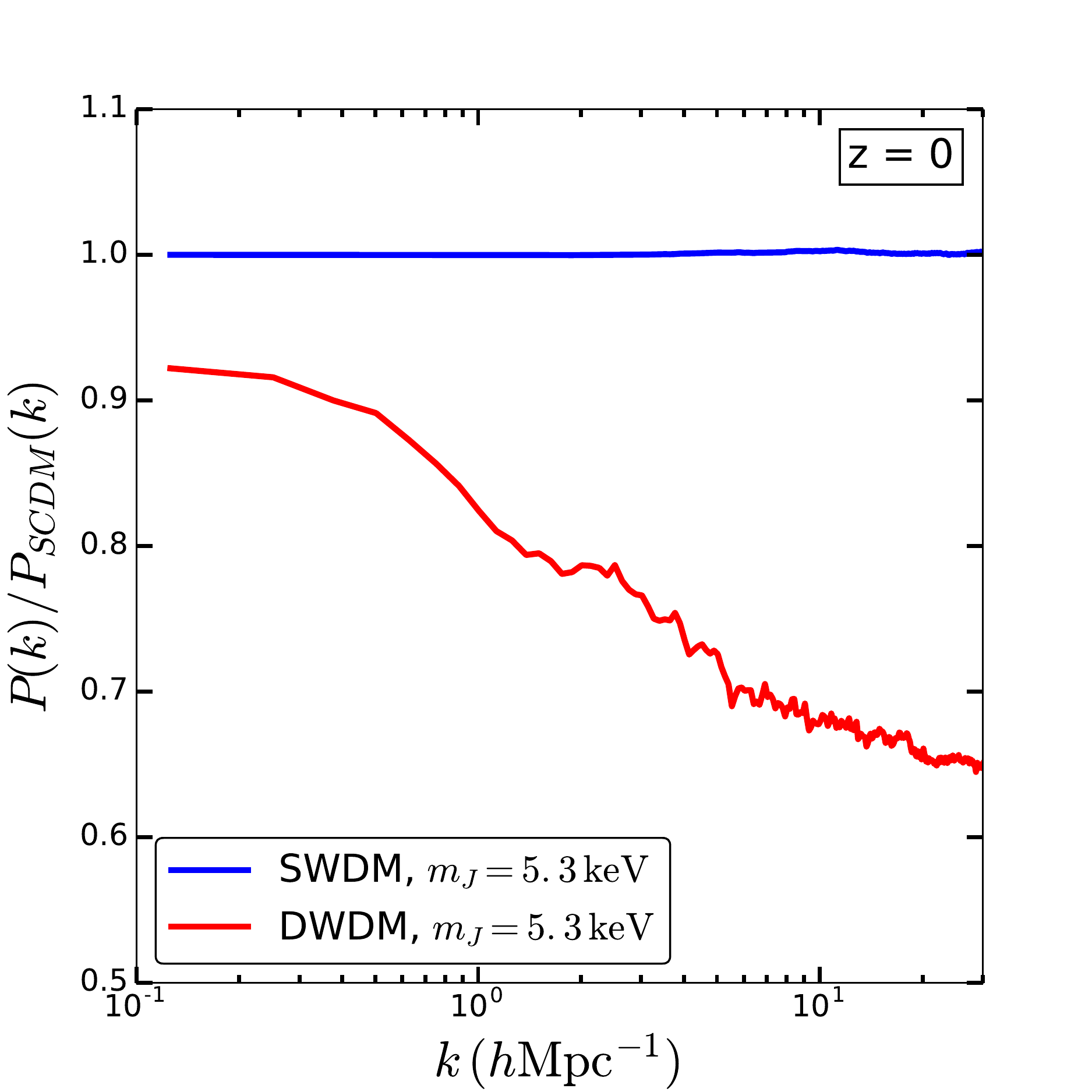}
\caption{Left panel: The matter power spectrum at $z=0$ for SCDM (black dashed),
 SWDM (blue solid) and DWDM (red solid) with $m_J = 5.3\,\mathrm{keV}$. 
 Right panel: Ratio between SWDM (blue solid) and DWDM (red solid) with 
 $m_J = 5.3\,\mathrm{keV}$ and SCDM.
 The ratio between SWDM and
 SCDM is very close to $1$ on all scales, due to the small free-streaming length
 of such a heavy WDM particle.
 }
\label{Fig:Pk_compare_Lyman}
\end{figure}

\begin{figure}[b]
\centering
\includegraphics[width=0.45\textwidth]{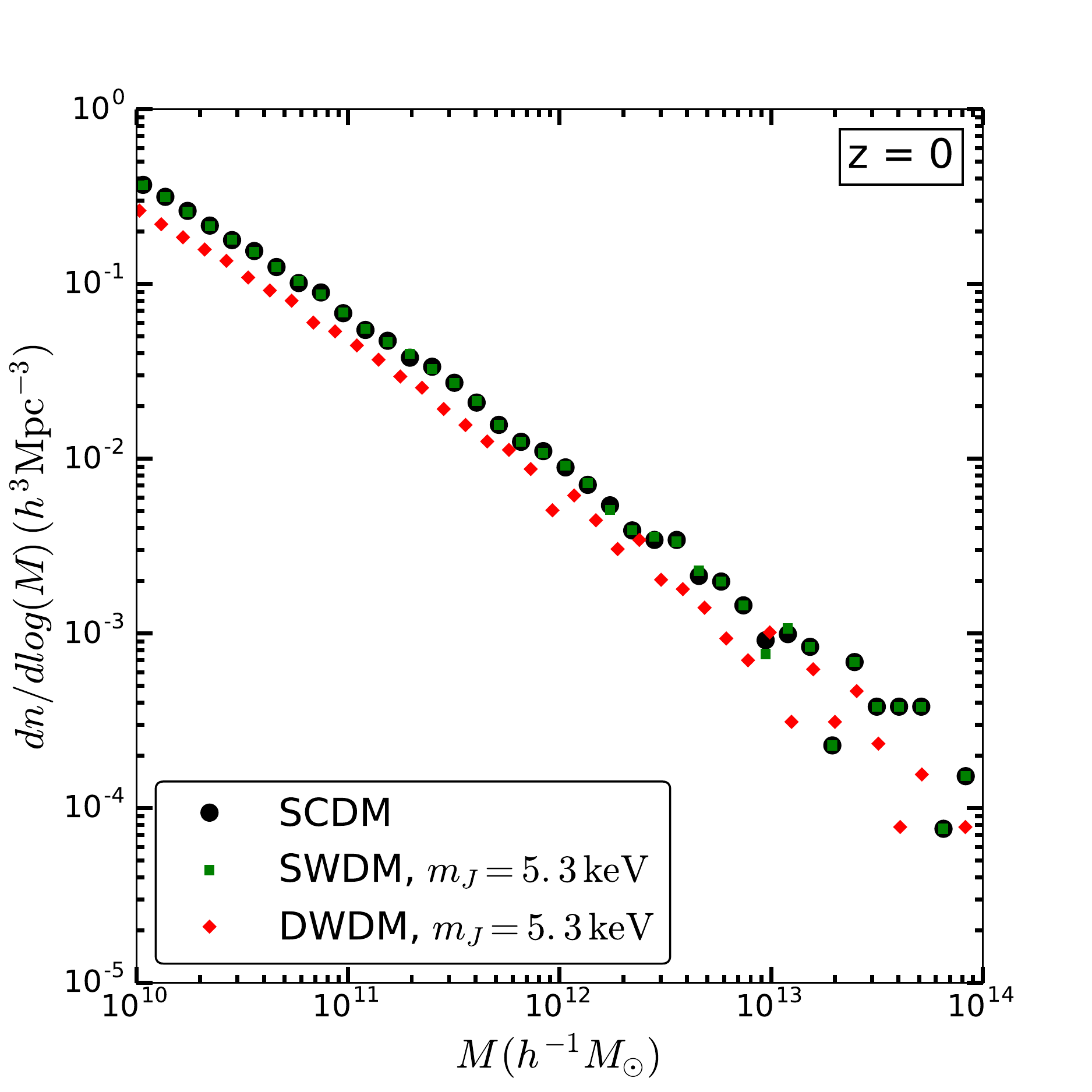}
\caption{The halo mass function at $z=0$ for SCDM (black circle), SWDM (green square) 
and DWDM (red diamond) with $m_J = 5.3\,\mathrm{keV}$. Like the matter power spectrum, 
the halo mass function is similar for SCDM and SWDM with $m_J = 5.3\,\mathrm{keV}$, despite
for some deviations in the high-mass end related to cosmic variance.
The halo mass function of DWDM with $m_J = 5.3\,\mathrm{keV}$ shows suppression of the halo number
density compared to that of SCDM and SWDM at all mass scales, as discussed in Sec.~\ref{Subsec:HMF}.
}
\label{Fig:HMF_Lyman}
\end{figure}

Although the uncertainty in the evolution of the IGM temperature might cast doubt on the interpretation of the Lyman-alpha forest~\cite{Hui:2016ltb,Zhang:2017chj} measurements, we note that recent Lyman-alpha forest observations may set a strong lower limit on the WDM mass.
Therefore, for completeness, we also perform simulations using the 95\% CL lower limit on the mass of the WDM particle allowed by Lyman-alpha forest~\cite{Irsic:2017ixq} data, i.e.  $m_{J} \geq 5.3\,\mathrm{keV}$. We keep the lifetime $\tau_J = 50\,\mathrm{Gyr}$ as in the other simulations with decay. 
In this appendix, we present the results with such a mass for both stable and decaying dark matter.

In Fig.~\ref{Fig:Pk_compare_Lyman}, we compare the matter power spectrum of SWDM and DWDM with $m_J = 5.3\,\mathrm{keV}$ to that of SCDM. 
We show the individual matter power spectra at $z=0$ in the left panel, and the ratios to the SCDM matter power spectrum in the right panel.
Note that the difference between SWDM with $m_J = 5.3\,\mathrm{keV}$ and SCDM is smaller than $1\%$ on all scales. This is associated to the relatively small free-streaming length of such a ``large'' mass WDM particle. 
Furthermore, a visual comparison with the red solid curve in the left panel of Fig.~\ref{Fig:Pk_ratiocompare}, shows that the power suppression due to the decay 
is in practice the same for WDM with $m_J = 5.3\,\mathrm{keV}$ and DCDM. This is again a consequence of the small free-streaming length of the WDM.

Similar considerations apply to the halo mass functions for SCDM, SWDM and DWDM, shown in Fig.~\ref{Fig:HMF_Lyman}. 
The number densities of halos are almost identical for SCDM and SWDM with $m_J = 5.3\,\mathrm{keV}$, except for some deviations in the high-mass end due to cosmic variance. 
Also note that the large number of spurious halos that were seen in the light WDM simulations discussed in Sec.~\ref{Subsec:HMF} disappear for WDM with $m_J = 5.3\,\mathrm{keV}$.
Moreover, the decay suppresses the halo mass function of DWDM on all scales.

From the analysis of the matter power spectrum and the halo mass function, we conclude that the WDM mass allowed by the Lyman-alpha forest is, at the scales probed by our analysis, undistinguishable from CDM. This holds for both the stable and decaying case.

\newpage

\black
\bibliographystyle{utphys}
\bibliography{bibliography}
\end{document}